\documentclass[twocolumn,aps,showpacs,longtable]{revtex4}

\usepackage{amsfonts,amsmath,graphicx,latexsym}
\usepackage{epsfig}

\newcommand{\vc}[1]{\mathbf{#1}}
\newcommand{\g}{\frac{4\pi a}{m}}

\begin{document}
\title{Collisionless dynamics of the condensate 
predicted in the random phase approximation}
\author{Patrick Navez}
\affiliation{
Ecole Polytechnique, CP 165, Universit\'e Libre de Bruxelles, 1050
Brussels, Belgium}

\date{\today}
\begin{abstract}
From the microscopic theory, we derive a number conserving
quantum kinetic equation, valid for a dilute Bose  gas
at any temperature,
in which the binary collisions between the quasi-particles are
mediated by phonon-like
excitations (called ``condenson''). This different approach
starts from the many-body Hamiltonian of a Boson gas
and uses, in an appropriate way, the generalized random phase 
approximation.
As a result, the collision term of the kinetic equation  
contains higher order contributions in the expansion in the 
interaction parameter. This different expansion 
shows up that a scattering involves the 
emission and the absorption of a phonon-like excitation.
The major interest of this particular mechanism is 
that, in a regime where the condensate is stable,
the collision process between condensed and non condensed particles 
is totally blocked due to a total annihilation 
of the mutual interaction potential induced by the condensate itself. 
As a consequence, the condensate is not constrained 
to relax and can be superfluid.
Furthermore, a Boltzmann-like 
H-theorem for the entropy exists for this equation and
allows to distinguish between
dissipative  and non dissipative phenomena (like vortices).
We also illustrate the analogy between this approach and
the kinetic theory for a plasma, in which the excitations correspond
precisely to a plasmon. Finally, we show the equivalence of this 
theory with the non-number conserving Bogoliubov theory 
at zero temperature.
\end{abstract}

\maketitle

\section{Introduction}

\subsection{Superfluidity and H-theorem}

A lot of studies have been devoted to the theoretical 
understanding of statistical and dynamical properties 
of a weakly interacting Bose condensed gas. In particular, 
many works have been accomplished on the derivation of quantum 
kinetic equations (QKE) that govern the evolution of the condensate 
fraction together with his thermal excitations 
\cite{Zaremba,Walser,Kirkpatrick,HM,LevichYakhot,Stoof,
Gardiner,Stoof2,Pomeau,Khalatnikov,Proukakis}. 
In a simple microscopic model of an homogeneous gas, 
one can describe the condensate by the atoms 
that populate the lowest ground state of energy and the thermal 
excitations by the atoms contained in the 
excited energy levels. The population of atoms in each level evolves 
according to the probability of scattering between the atoms. In 
the case of a uniform gas, the so-called Uehling-Uhlenbeck quantum 
kinetic equation (UUQKE) is a 
non-linear integral equation which describes 
the detailed balance of  
the population transfer of atoms for each mode of the wave-vector through 
a binary collision term \cite{Balescu}. 
This term depends linearly on the scattering differential 
cross section and nonlinearly on the mode population. It has also 
the remarkable 
properties to allow the QKE to obey conservation laws. 
These laws guarantee that the total number, the total momentum and the 
total kinetic energy 
of particles are preserved during the collision processes. More striking 
is the law stating that the production of 
entropy must be always positive, 
guaranteeing that the system obeys the second law of 
thermodynamics and, consequently, 
that it is always dissipative. 
This important requirement is known as the Boltzmann 
H-theorem established for a classical gas. 

The UUQKE has been derived in a weak coupling approximation, valid 
for a diluted gas for which the kinetic energy is much higher 
than the potential energy. The collision term is indeed a second order 
expansion in the interaction potential between atoms leading to 
an  expression of the differential cross section in the Born 
approximation. 

One could use such a QKE in a regime below 
the critical point of condensation. In particular, for an 
inhomogeneous
gas in a trap potential, 
the kinetic equation describes the evolution of the 
Wigner function and must contain additional terms,
taking into account  
the free propagation of the atoms, the 
influence of the external potential and the Hartree-Fock mean field.
The conservation laws are still valid and allow to express the 
hydrodynamic equations, including the equation for the local entropy 
production. 

Despite these consistencies, the resulting QKE suffers from the 
lack of understanding of an important phenomenon: {\it superfluidity} 
\cite{Leggett2}. 
If this phenomenon really exists for a dilute gas \cite{Cornell}, 
then the second order 
theory must necessarily be revisited as it does not take into account the 
frictionless motion. Indeed according to the H-theorem, the homogeneous 
gas with a zero total momentum evolves towards a statistical 
equilibrium state characterized by the Bose-Einstein particle number 
distribution $n^{eq}_{\vc{k}}=1/[\exp(\beta(\epsilon_{\vc{k}}-\mu))-1]$ 
where $\vc{k}$ is the wave-vector, 
$\beta=1/k_B T$ the inverse temperature, $\mu$ the chemical potential, 
and $\epsilon_{\vc{k}}=\vc{k}^2/2m$ the particle kinetic energy 
respectively ($\hbar=1$). 
For $\mu \rightarrow 0$ and $\vc{k} \rightarrow 0$, the ground state 
has a macroscopic population $n_{\vc{0}}= 1/[\exp(-\beta\mu)-1]$ having no 
relative velocity with the non condensed part of the gas.  A non zero 
relative velocity corresponds to a non equilibrium situation and collisions 
between condensed and non condensed atoms will irrevocably damp 
this velocity towards zero. 

From this observation, we conclude that the second order theory is 
no longer valid as far as superfluidity is concerned. Obviously, 
at zero temperature, only the potential energy is the 
dominant contribution since 
the condensate is at rest and no thermal excitation subsists. 
Moreover, other indications confirm that an improved description of 
a weakly interacting Bose gas requires a higher order analysis in the 
interaction parameter.
Among them, let us 
mention that the entropy of the condensate must be zero or close to zero 
while
the H-theorem predicts an entropy $S_0 \sim \log n_{\vc{0}}$ corresponding to 
a system in the Grand Canonical ensemble with large statistical 
fluctuations of the particle number compared to his average value 
$\delta n_{\vc{0}}/ n_{\vc{0}} \sim 1$ \cite{Huang}. This is due to the 
Bose enhancement factor which stimulates the collision rate in 
a huge manner as long as a condensed particle is involved in the process. 
On the other hand, the equilibrium statistical approach based 
on the partition 
function formalism indicates that  
the presence of a small interaction lowers considerably 
the fluctuations to an amount irrelevant in the thermodynamic 
limit $\delta n_{\vc{0}}/ n_{\vc{0}} \rightarrow 0$ \cite{4eme}. 

Landau was the first to give an explanation of  
the superfluidity mechanism. He found 
a necessary but not sufficient condition for which this 
phenomenon happens \cite{LL,HM}. He showed that a 
superfluid interacting with an external body 
(for example the wall) cannot release its 
energy, unless it evolves with a velocity higher 
than a critical one. The argument is based 
on the impossibility to satisfy the momentum-energy conservation 
requirement because the excitation emitted by the superfluid 
has a phonon-like dispersion relation.  
  
\subsection{Beyond the second order perturbation theory}

Attempts to analyze contribution coming from higher terms have been 
carried out with some success ( see \cite{Leggett,Pines} for a review). 
At zero temperature, Bogoliubov 
has calculated corrections to the ground state energy. 
The result is 
non analytic in the interaction parameter, due to infrared 
divergencies, and is obtained through a re-summation of an infinite 
number of contributions. Moreover, the theory 
predicts the existence of a phonon-like 
excitation, necessary in order to justify the Landau mechanism for 
superfluidity. 

Using Green function techniques, Beliaev improved 
the description of the phonon like excitation namely 
by calculating its damping rate. Later on, Hugenholtz 
and Pines (HP) demonstrated that this 
excitation must necessarily be gapless.
If at zero temperature, the theory seems well 
understood and widely accepted, for finite 
temperature it seems rather controversial.
Popov was one of the first to extent the Bogoliubov 
and Beliaev works using the Matsubara formalism 
at finite temperature \cite{Popov}. However, this 
approach is essentially valid for a weakly depleted 
gas and thus for temperatures much below 
the critical one $T_c$. To take into account a 
strong depletion, Girardeau derived the 
Hartree-Fock-Bogoliubov (HFB) mean field equations which 
unfortunately, have, the strong inconvenient of 
having a gap in the quasi-particle energy  spectrum \cite{Girardeau}. 

Since then, 
many attempts, have been made, in order 
to suppress this gap and ``rescue'' the HFB equations, with 
the purpose 
to go beyond the Bogoliubov phonon-like
dispersion relation.
\cite{Griffin,Gapless,PN}. 
A very popular one is the so-called 
"Popov approximation" which consists in suppressing 
the product of anomalous term in the HFB equations \cite{Griffin}. 
Another consists in the renormalization of the HFB equations 
but at the price of making the approximation
of a weakly depleted Bose gas
\cite{Gapless}.
Consequently, these approaches are only strictly valid for temperatures 
much below $T_c$.
{\it A derivation of a well-defined theory, 
which is able to describe the weakly interacting Bose  
gas for any regime of temperature, is still an open 
problem and remains to be established} (see the conclusion of 
\cite{NP2}). By well-defined, 
we mean that the expansion in the small interaction 
parameter must be valid, whether the condensate is strongly 
depleted or not.

Nevertheless, using these approaches,  
QKE have been derived taking into account these higher order 
effects in the collision term. 
While some authors use the dispersion relation  
resulting from the Bogoliubov theory \cite{Gardiner,Kirkpatrick,Imamovic}, 
others use the 
one  resulting form the renormalized HFB theory \cite{Walser}. 
Superfluidity can be achieved in these models. Indeed, 
the resulting QKE can preserve detailed balance even 
for metastable states, in which a non zero relative 
velocity persists. In this situation, the Bogoliubov 
dispersion relation becomes asymmetric - due to a Lagrange 
multiplier representing the relative velocity - and 
remains non negative, as long as this velocity does not exceed 
the sound velocity. Otherwise, the dispersion relation 
is negative and the gas becomes unstable \cite{Huang}.   
Although these models represent a 
considerable amount of
work, they do not yet provide a full account of 
the conservation laws and the H-theorem 
that might be deduced from their QKE. 

However, let us mention that a QKE has been proposed 
valid in principle for temperatures close to $T_c$ \cite{Zaremba, Pomeau}. 
The main problem of this approach is that the dispersion 
relation has a gap and thus fails to explain the superfluidity. 

In general, in all these models, 
collisions between condensed and non 
condensed atoms are possible and their rate is 
huge because of the Bose enhancement factor in the same way 
as in the UUQKE. If we follow the same reasoning as above and 
if an H-theorem really exists, we might expect a  
condensate entropy of the same order of magnitude 
$S_0 \sim \log n_{\vc{0}}$.  
As a consequence, 
the particle number fluctuations  of the 
condensed mode are also huge, in contradiction with 
estimations made from the partition function formalism 
\cite{4eme}.

\subsection{The random phase approximation approach}

The present paper is devoted to provide a 
possible alternative explanation to the problem of superfluidity 
in a diluted Bose gas. 
For this purpose a different QKE is derived in the 
random
phase approximation (RPA) in   
which higher order terms in the interaction parameter 
are retained in the binary collision term (in some 
work, the RPA QKE refers to the UUQKE and thus has a different 
meaning from this paper \cite{LevichYakhot}).
{\it 
Furthermore, in comparison with other approaches, 
the QKE is valid for any regime of temperature below 
and above $T_c$ since, under no circumstances, the 
approximation of weak depletion has been used.}   

The RPA is commonly used to describe the 
collective excitations in
the quantum plasma of an electron liquid \cite{NP}. 
The idea behind this approximation is to neglect  
contribution containing 
average over pair of field operators that are not oscillating 
in phase but rather randomly.
In 
the terminology of optics, we neglect contribution that are not 
phase matching. 
These contributions come essentially from averages over  
pairs of different modes which oscillate with 
a relative random phase.
The reason for using the RPA in a diluted  
plasma is 
that, we expect that the excitations propagate over a 
sufficiently long time 
that non phase-matching terms are destroyed by 
interference. By analogy with optics, we make somehow the far field 
approximation. The analysis of collective excitations in plasma 
using RPA reveals that 
the coulombian interaction potential between the electrons 
is shielded at long distance, due to the dynamic dielectric 
function. In  fact, these results generalize 
the Debye theory predicting the screening of the potential 
between a charged ion due to the presence of other   
surrounding ions.
As an extension, the RPA includes also the dynamical aspects of 
these charged particles. 

A QKE for the plasma has been 
derived \cite{QBalescu,WP}. It predicts that 
the coulombian interaction potential 
for the cross section is shielded precisely by this dynamic 
dielectric factor. 
The classical version of this equation is known 
as the BGL kinetic equation 
(Balescu-Guernsey-Lenard) \cite{Balescu}. For deriving it, Balescu uses 
a re-summation of the so-called ring diagrams, which is another 
approach equivalent to the RPA \cite{QBalescu}. Later on, 
Wyld and Pines 
established a connection between the QKE and the plasmon theory \cite{WP}. 
In their approach, the  
shielded potential results from a more subtile dynamical 
mechanism, in 
which the two  electrons must emit and reabsorb an 
intermediate plasmon during their interaction. 
The dispersion relation of the plasmon corresponds 
precisely to the collective mode of the plasma and 
its decay rate to the Landau damping.

In this paper, an identical type of approach will be reproduced in the 
case of a dilute Bose gas. It is already 
known that the RPA allows to recover the Bogoliubov 
results for the ground state energy but with one 
important difference \cite{Pines}. Namely, in the RPA,
there is no  need of a spontaneous breaking of the symmetry $U(1)$ 
but rather, during the derivation, the total particle number 
is kept conserved. 
Note that a number-conserving formalism has been already 
used to derive a QKE \cite{Gardiner}. But this formalism  is 
based on a $1/N$ expansion \cite{GCD} and thus differs from 
the one in the RPA \cite{Pines}.

These considerations 
allow to derive a number conserving QKE in which the 
interaction potential is also modified by the presence 
of a dynamic dielectric function resulting from collective 
excitations. However, in contrast to a plasma,  
this function presents a strange behavior 
according to the kind of quasi-particles interaction it 
mediates. {\it When the interaction involves a particle of the 
macroscopic condensate,
it has the effect to totally annihilate the potential, forbidding 
any binary collision to occur.} 
 
This surprising result allows to explain why a 
metastable condensate cannot decay into a state of 
lower energy. Simply, it is forbidden for a 
particle of the condensate to scatter with the 
excited particle and to become thus excited. 
No exchange of particle can occur between the two 
fluids. This unexpected phenomenon is a consequence 
of the collective behavior induced by the presence 
of a macroscopic condensate. In short, collision cannot 
happen because an induced 
mean field force, generated by the condensate, 
compensates exactly the interaction 
force felt by the condensed and non condensed particle. 
The net result is the absence of an effective interaction 
force preventing any scattering.  
In other words, the dielectric function used to attenuate the 
binary interaction potential becomes simply infinite.


This phenomenon of ``collision blockade'' also influences 
the mechanism of condensate 
formation. Since particle cannot be exchanged with 
the macroscopic condensate, other mechanisms 
must be found. A closer analysis reveals that the collision 
blockade happens only when the Bose gas is stable. By stable we mean 
that the collective oscillations are always damped. Precisely, 
the Landau damping plays this role in both a plasma and a 
condensate as long as we are close to equilibrium \cite{Balescu,SK}. 
However, for some non equilibrium situation, it has been 
shown for a plasma that the oscillations can grow exponentially 
leading to an instability \cite{instB,FR}. Such behavior happens also for 
a Bose gas which becomes thus unstable. For example, we 
will show that this is the case when the relative velocity between 
the normal and superfluid is higher than the speed of sound. 
In such instable regime, the picture of a collision blockade 
is no longer valid and the QKE gets more 
complicated. Such more sophisticated QKE can describe the 
exchange of particle with the condensate \cite{instB,FR} and provides an 
alternative explanation for the condensate formation 
\cite{Keterlee2,BZS,Gardiner2}. 
Another possibility is to exchange 
them on the edge region where the condensate is not macroscopically 
populated so that scattering can occur. But this requires 
also a really specific analysis. Therefore, the present 
QKE derived here will not address the important issue 
of particle exchange with the condensate.  

Assuming that collision processes are local
in space, the derivation can be extended for 
a weakly inhomogeneous
Bose gas. 
In this way, we recover the generalized Gross-Pitaevskii
equation for the condensate, but with the difference that 
the absence of a binary collision term in this equation has 
now a clear justification. When the normal cloud is in a local 
thermal 
equilibrium, no dissipation exists anymore and 
we can derive a set of coupled equations 
defining 
the superfluid regime for any finite temperature below $T_c$.

The paper is divided as follows. 
We first begin by a heuristic approach of the ``collision blockade'' 
mechanism in section 2.
In section 3, we derive 
the QKE for a homogeneous Bose gas. We review the simple RPA in which 
only the Hartree or direct term is considered and 
the generalized RPA (GRPA) in which both Hartree and exchange Fock terms are 
considered \cite{NP}. An equation of motion for the particle-hole pair operator 
is derived from which the collective and scattering excitation 
energy spectra are deduced. The solution of the 
equation of motion allows to calculate the collision term and 
to predict the collision blockade phenomenon. We show 
that the QKE conserves the total particle number, the total momentum
and the total energy (kinetic and Hartree-Fock) 
and that a Boltzmann-like H-theorem exists. 
In section 4, the reasoning of the previous section is extended 
to the case of a weakly inhomogeneous Bose gas. From the resulting 
QKE, we delimitate the superfluid regime i.e. the regime  
in which there is no local production of entropy. In section 
5, by analogy with plasmon theory, 
we interpret the collision phenomenon as the result 
of the exchange of an intermediate boson (condenson). 
In section 6, we show how the number conserving RPA theory 
allows to recover the Bogoliubov results for the 
ground state energy.  Finally, section 7 is devoted 
to conclusions and perspectives

\section{The ``collision blockade'' phenomenon}

\subsection{Preliminary definitions}

We start from the Hamiltonian:
\begin{eqnarray}
H=\sum_{\vc{k}}
\epsilon_{\vc{k}}
c^\dagger_{\vc{k}}
c_{\vc{k}}
+\sum_{\vc{k},\vc{k'},\vc{q}}
\frac{U_\vc{q}}{2V}
c^\dagger_{\vc{k}+\vc{q}}c^\dagger_{\vc{k'}-\vc{q}}
c_{\vc{k}}c_{\vc{k'}}
\end{eqnarray}
$c^\dagger_{\vc{k}}$ and $c_\vc{k}$ are the creation and
annihilation operators obeying the commutation relations
$[c_\vc{k},c^\dagger_\vc{k'}]=\delta_{\vc{k},\vc{k'}}$
$[c_\vc{k},c_\vc{k'}]=[c_\vc{k}^\dagger,c^\dagger_\vc{k'}]=0$.
$\epsilon_{\vc{k}}=\frac{\vc{k}^2}{2m}$ is the kinetic
energy of the particle.
$U_{\vc{q}}$ is the interaction potential expressed in
Fourier transform.
Since we are concerned only with low energy
binary collisions in the channel $l=0$,
it is common to replace the potential $U_{\vc{q}}$
by a pseudo-potential or contact potential
that we treat in the Born
approximation \cite{LL}:
\begin{eqnarray}\label{U0}
U_{\vc{q}}=\frac{4\pi a}{m}
[1+\frac{4\pi a}{mV}\sum_{\vc{q'}}
\frac{1}{\epsilon_{\vc{k}+\vc{q}}
+\epsilon_{\vc{k'}-\vc{q}}
-\epsilon_{\vc{k}}-\epsilon_{\vc{k'}}}]
\end{eqnarray}
where $a$ is the scattering length.
Usually we consider only the first linear term in the
scattering length. The second term is ultra-violet divergent and
is only present to renormalize the
theory by  removing eventual high energy divergencies.
In what follows we will concentrate on a repulsive
contact interaction $a > 0$.

The so-called annihilation and creation operators of the particle-hole pair
are of interest:
\begin{eqnarray}
\rho_{\vc{k},\vc{q}}=c^\dagger_{\vc{k}}c_{\vc{k}+\vc{q}} 
\ \ \ \ \ 
\rho^\dagger_{\vc{k},\vc{q}}=c^\dagger_{\vc{k}+\vc{q}}c_{\vc{k}}
\end{eqnarray}
They represent an excitation of momentum $\vc{q}$ created from
a particle which transfers its momentum from $\vc{k}$ to
$\vc{k}+\vc{q}$. The kinetic energy transfer for 
this excitation is given by 
$\omega_{\vc{k},\vc{q}}= \epsilon_{\vc{k}+\vc{q}}
-\epsilon_{\vc{k}}$.
In particular, we define
the density fluctuation operator
$\rho_{\vc{q}}=\sum_{\vc{k}} \rho_{\vc{k},\vc{q}}=
\rho^\dagger_{\vc{-q}}$,
the number operator $\hat n_{\vc{k}}=\rho_{\vc{k},0}$ 
and the total number $\hat N = \sum_{\vc{k}}
\hat n_{\vc{k}}$.
In terms of these operators, the Hamiltonian becomes:
\begin{eqnarray}
H=\sum_{\vc{k}}
\frac{\vc{k}^2}{2m}
\hat n_{\vc{k}}
+\sum_{\vc{q}}
\frac{U_{\vc{q}}}{2V}
(\rho^\dagger_{\vc{q}}\rho_{\vc{q}}-
\hat N)
\end{eqnarray}

\subsection{Heuristic approach}

Assume a macroscopic condensate containing 
$n_{\vc{k_s}}$ particle evolving with momentum $\vc{k_s}$. 
In the absence of excitations, there are no fluctuations 
of the density operator i.e. 
$\langle \rho_\vc{q}\rangle^{eq}=
\delta_{\vc{q},\vc{0}}n_{\vc{k_s}}$.
Suppose that an external potential is turned 
on creating locally fluctuations of 
the density of the condensate. Expressing 
these perturbations in Fourier space, we can 
characterize the external 
potential $\phi_{ext}(\vc{q},\omega)$ and 
the density fluctuations $\delta n(\vc{q},\omega)=
\langle \delta \rho_\vc{q}\rangle=
\langle \rho_\vc{q} \rangle-
\langle \rho_\vc{q} \rangle^{eq}$ 
in terms of its wave-vector $\vc{q}$ and frequencies 
$\omega$ components.
The Hamiltonian created 
by this external potential is given by 
$H_{ext}(t)=\sum_{\vc{q}}
\rho_{\vc{q}}^\dagger e^{-i\omega t}\phi_{ext}(\vc{q},\omega)
+ c.c.$
The linear response to this potential is 
given by the formula:
\begin{eqnarray}\label{resp1}
\delta n(\vc{q},\omega)
=\chi (\vc{q},\omega)
\phi_{ext}(\vc{q},\omega)
\end{eqnarray} 
where $\chi (\vc{q},\omega)$ is the susceptibility 
function. This response function can be calculated formally 
by treating $H_{ext}(t)$ as a perturbation in the first order. 
Without the presence of 
a binary interaction potential between the particle, 
this function is given by
\cite{NP}:
\begin{eqnarray}
\chi_0 (\vc{q},\omega)=
\frac{n_{\vc{k_s}}}{
\omega-\omega_{\vc{k_s},\vc{q}}+i0_+}
-
\frac{n_{\vc{k_s}}}{
\omega+\omega_{\vc{k_s-q},\vc{q}}+i0_+}
\end{eqnarray} 
It represents the transition amplitude
for a condensed particle
to increase its kinetic energy by an amount
$\omega_{\vc{k_s},\vc{q}}=\epsilon_{\vc{k_s}+\vc{q}}
-\epsilon_{\vc{k_s}}$  minus the amplitude
for an excited particle to be transferred to
the condensate releasing the energy
$-\omega_{\vc{k_s-q},\vc{q}}=\epsilon_{\vc{k_s}-\vc{q}}
-\epsilon_{\vc{k_s}}$.
An infinitesimal quantity $i0_+$ has been 
added to ensure the convergence and is due to 
the adiabatic switching process of external perturbation.
In the presence of the interaction, the 
condensate acquires a self-interaction 
potential energy 
given by $U_{\vc{0}}n_{\vc{k_s}}^2/(2V)$. 
Moreover, according to the RPA, 
the presence of fluctuation 
density changes the potential energy by 
the amount $\delta H_{int}=
\sum_{\vc{q}}
\rho_{\vc{q}}^\dagger 
e^{-i\omega t}\delta \phi(\vc{q},\omega) + c.c.$
where $\delta \phi(\vc{q},\omega)=(U_{\vc{q}}/V)
\langle \delta \rho_{\vc{q}} \rangle$ 
is the
potential induced by the presence of 
the density fluctuations.  
Consequently, the global response of the 
system becomes:
\begin{eqnarray}\label{resp2}
\delta n(\vc{q},\omega)
&=&\chi_0 (\vc{q},\omega)
\left(\phi_{ext}(\vc{q},\omega)+\delta \phi(\vc{q},\omega)\right)
\nonumber \\
&=&\chi_0 (\vc{q},\omega)
\left(\phi_{ext}(\vc{q},\omega)+(U_{\vc{q}}/V)\delta n(\vc{q},\omega)\right)
\end{eqnarray}
Comparison between $(\ref{resp1})$ and 
$(\ref{resp2})$ allows to deduce:
\begin{eqnarray}\label{susc}
\chi (\vc{q},\omega)=\frac{\chi_0 (\vc{q},\omega)}
{1- (U_{\vc{q}}/V)\chi_0 (\vc{q},\omega)}
\end{eqnarray}
The total potential created inside the system 
$\phi_{tot}(\vc{q},\omega)=
\phi_{ext}(\vc{q},\omega)+\delta \phi(\vc{q},\omega)$
is related to the external potential 
through the dynamical dielectric function:
\begin{eqnarray}\label{heur}
{\tilde {\cal K}}(\vc{q},\omega)=
\frac{\phi_{ext}(\vc{q},\omega)}{
\phi_{tot}(\vc{q},\omega)}=
1 - \frac{U_{\vc{q}}}{V}\chi_0 (\vc{q},\omega)
\end{eqnarray}
In the electromagnetism language, the gradient of 
$\phi_{ext}(\vc{q},\omega)$ corresponds to the electrical 
displacement, 
while the gradient of $\phi_{tot}(\vc{q},\omega)$ corresponds 
to the electric field. 
These are respectively  the external and 
total field force acting on the particle. 
The dielectric function usually has the effect of attenuating 
the external field force by means of an induced force so that 
the total field force is smoothed out. In particular,
we notice that this function is infinitely 
resonant for frequencies $\omega=\omega_{\vc{k_s},\vc{q}} , 
-\omega_{\vc{k_s}-\vc{q},\vc{q}}$ 
which means that the induced potential exactly 
compensates the external one. For these frequencies, 
the external potential affects the density fluctuations but 
has no local influence anymore on the condensate particle itself. 
This important observation is at the origin of the collision 
blockade phenomenon.

Indeed, assume that the source of 
this finite external potential results from 
the transition of an excited particle 
of momentum $\vc{k'}$ towards another 
excited state with a momentum 
$\vc{k'}-\vc{q}$. The  
released energy during this process is 
$\omega=-\omega_{\vc{k'},-\vc{q}}$.
According to the Fermi golden rule, a collision 
occurs between the condensed and non 
condensed ingoing particles if the transfer energy  
are equal $-\omega_{\vc{k'},-\vc{q}}=\omega_{\vc{k_s},\vc{q}}$ 
or equivalently the total energy is conserved  
$\epsilon_{\vc{k_s}}+\epsilon_{\vc{k'}}=
\epsilon_{\vc{k_s}+\vc{q}}+\epsilon_{\vc{k'}-\vc{q}}$. 
But then, the dielectric function 
becomes infinitely resonant ${\tilde {\cal K}}(\vc{q},\omega) 
\rightarrow \infty$ causing  
the total potential 
$\phi_{tot}(\vc{q},\omega) \rightarrow 0$. Thus 
the interaction potential felt by the condensate 
particle becomes non-existent, since the induced 
mean field cancels the external potential generated 
by the excited particle. 
That precise 
phenomenon is responsible for the 
absence of an effective  scattering process since, 
without interaction potential, no scattering amplitude 
can appear. The same effect happens when the scattering  
involves an outgoing particle in the condensate whereas, 
in such a case, the transfer energy  
$-\omega_{\vc{k_s-q},\vc{q}}$ causes the infinite resonance.
Thus, these giant resonances have the effect to protect 
the condensate particle from scattering with the others.
We must notice that this mechanism works only 
if the condensate population is macroscopic, 
otherwise, no induced potential is generated  
in the thermodynamic limit since, for 
$n_{\vc{k_s}}/V \rightarrow 0$, ${\tilde {\cal K}}(\vc{q},\omega)
\rightarrow 1$.
As we shall see in the next sections, 
a more elaborated model confirms this 
prediction in the more general case 
of a non equilibrium Bose gas, in which 
the condensate can be strongly 
depleted.

\section{Kinetic theory in the RPA}

Many methods have been developed to derive kinetic equations 
for a dilute Bose gas
\cite{Walser,Imamovic}.
In order to arrive rapidly to the final result , instead of using
complicated many body techniques, we base the derivation on an
operatorial method developed in \cite{NP} which appeared to be
a simpler way to reach results without loss of generality.
The approximations are the followings:
1) homogeneous Bose gas, 2) thermodynamic limit, 3) generalized RPA,
4) instantaneous  collisions (Markovian QKE), 5) no fragmentation
of the condensate (only one macroscopic population mode).

\subsection{The random phase approximation}

In this subsection, we give a brief overview 
of the RPA developed in \cite{NP}.
This approximation has been used quite extensively 
for the quantum electron liquid for describing 
the screening effect. 

The dynamic of $\rho_{\vc{k},\vc{q}}$ is given by 
the Heisenberg equation motion. Using the relation: 
\begin{eqnarray}
[\rho_{\vc{k},\vc{q}},\rho_{\vc{k'},\vc{q'}}]
=\delta_{\vc{k}+\vc{q},\vc{k'}}
\rho_{\vc{k},\vc{k'}+\vc{q'}-\vc{k}}
-\delta_{\vc{k}-\vc{q'},\vc{k'}}
\rho_{\vc{k'},\vc{k}+\vc{q}-\vc{k'}}
\end{eqnarray}
we find
\begin{eqnarray}\label{he}
i\frac{\partial}{\partial t}\rho_{\vc{k},\vc{q}}
=[\rho_{\vc{k},\vc{q}},H]
=(\epsilon_{\vc{k}+\vc{q}}
-\epsilon_{\vc{k}})
\rho_{\vc{k},\vc{q}}
\nonumber \\
+\sum_{\vc{q'}}\frac{U_{\vc{q'}}}{2V}
[\rho_{\vc{k},\vc{q}+\vc{q'}}
-\rho_{\vc{k}-\vc{q'},\vc{q}+\vc{q'}},
\rho^\dagger_{\vc{q'}}]_+
\end{eqnarray}
where the brackets refer to 
to an anti-commutator.
Two cases in the RPA are generally considered 
in systems close to equilibrium \cite{NP}. 
The simple RPA, in which only the Hartree or direct terms 
contribute, and the generalized RPA, in which the Fock 
or exchange terms are also retained.
We shall concentrate on the second approximation, since for 
a contact potential Hartree and Fock terms are identical.
Only for the condensate-condensate interaction, 
the Fock term does not appear. In the GRPA, 
we consider that operator $\rho_{\vc{k},\vc{q}}$
with non zero momentum transfer $\vc{q}\not=0$ 
gives a negligible contribution in comparison to 
$\hat n_{\vc{k}}$, as it involves different modes oscillating 
with a random relative phase.

The procedure to get the non equilibrium RPA equations is 
as follows. For a momentum transfer $\vc{q}=0$ the Eq.(\ref{he}) 
is kept unchanged. For $\vc{q}\not=0$, however, we keep among 
all terms those combinations of creation and annihilation 
operators involving product of $\rho_{\vc{k'},\vc{q}}$ 
and $\hat n_{\vc{k''}}$ for all possible values of $\vc{k'}$ 
and $\vc{k''}$, 
and neglect those combinations that cannot be written in this form.
These removed contributions are quadratic in 
the operator  
$\rho_{\vc{k'},\vc{q'}}$   
with $\vc{q'}\not=\vc{q},\vc{0}$. In this approximation, only 
contributions conserving the momentum transfer 
are relevant, the others coupling the various 
$\rho_{\vc{k},\vc{q}}$ with different 
momentum transfers are neglected. We expect that 
the 
excitations of momentum $\vc{q}$ propagate, without 
interfering with the others, over an 
enough long time determined by the dilution of the gas, 
that  (this would
correspond to the far field limit). 
The result is for $\vc{q}=0$
\begin{eqnarray}\label{n}
i\frac{\partial}{\partial t}\hat n_{\vc{k}}
=\sum_{\vc{q'}}\frac{U_{\vc{q'}}}{2V}
[\rho_{\vc{k},\vc{q'}}
-\rho_{\vc{k}-\vc{q'},\vc{q'}},
\rho^\dagger_{\vc{q'}}]_+
\end{eqnarray}
and for $\vc{q}\not=0$
\begin{eqnarray}\label{rho}
i\frac{\partial}{\partial t}\rho_{\vc{k},\vc{q}}
&=&
(\epsilon_{\vc{k}+\vc{q}} -\epsilon_{\vc{k}})
\rho_{\vc{k},\vc{q}}
\nonumber \\
&+&\sum_{\vc{k'}}[\frac{U_{\vc{k}-\vc{k'}+\vc{q}}}
{2V}\hat n_{\vc{k'}}- \frac{U_{\vc{k}-\vc{k'}-\vc{q}}}
{2V}\hat n_{\vc{k'+q}},\rho_{\vc{k},\vc{q}}]_+
\nonumber \\
&+&\sum_{\vc{k'}\not= \vc{k} }\frac{U_{\vc{q}}}{2V}
[\hat n_{\vc{k}}- \hat n_{\vc{k+q}},\rho_{\vc{k'},\vc{q}}]_+
\nonumber \\
&+&\sum_{\vc{k'}\not=\vc{k}-\vc{q}}\frac{U_{\vc{k}-\vc{k'}}}
{2V}[\hat n_{\vc{k}},\rho_{\vc{k'},\vc{q}}]_+
\nonumber \\
&-&\sum_{\vc{k'}\not=\vc{k}+\vc{q}}\frac{U_{\vc{k}-\vc{k'}}}
{2V}[\hat n_{\vc{k}+\vc{q}},\rho_{\vc{k'},\vc{q}}]_+
\end{eqnarray}
Eq.(\ref{rho}) is an integral operatorial equation 
linear in $\rho_{\vc{k},\vc{q}}$. We can linearize 
this equation by averaging all possible bilinear 
contributions. Since the homogeneity of the gas imposes
$\langle \rho_{\vc{k},\vc{q}} \rangle =0$ if 
$\vc{q}\not= \vc{0}$, we are left with 
the average on the number operator 
$\langle \hat n_{\vc{k}} \rangle= n_{\vc{k}}$ 
and we obtain 
an integral equation which possesses the same structure as Eq.(5.183)
p. 318 of \cite{NP}:
\begin{eqnarray}\label{rho3}
\left[i\frac{\partial}{\partial t}
-(\epsilon^{HFA}_{\vc{k}+\vc{q}}-\epsilon^{HFA}_{\vc{k}})\right]
\rho_{\vc{k},\vc{q}}=
(n_{\vc{k}}-n_{\vc{k+q}})
\sum_{\vc{k'}\not=\vc{k}}\frac{U_{\vc{q}}}{V}\rho_{\vc{k'},\vc{q}}
\nonumber \\
+n_{\vc{k}}
\sum_{\vc{k'}\not=\vc{k}-\vc{q}}\frac{U_{\vc{k}-\vc{k'}}}
{V}
\rho_{\vc{k'},\vc{q}}
-n_{\vc{k+q}}\sum_{\vc{k'}\not=\vc{k}+\vc{q}}\frac{U_{\vc{k}-\vc{k'}}}
{V}\rho_{\vc{k'},\vc{q}}
\nonumber \\
\end{eqnarray}
Indeed, in the term containing the bracket, 
we recognize the difference 
$\omega_{\vc{k},\vc{q}}=
\epsilon^{HFA}_{\vc{k}+\vc{q}}-
\epsilon^{HFA}_{\vc{k}}$
between the quasi-particle 
energy calculated in Hartree-Fock approximation (HFA): 
\begin{eqnarray}
\epsilon^{HFA}_{\vc{k}}=\frac{\vc{k}^2}{2m} 
+ \sum_{\vc{k'}}\frac{U_{\vc{0}}}{V}n_{\vc{k'}}+
\sum_{\vc{k'}}\frac{U_{\vc{k}-\vc{k'}}}{V}n_{\vc{k'}}
\end{eqnarray}
On the other hand, in the integral terms, 
both Hartree and Fock 
terms are present. However, Eq.(\ref{rho3}) shows some differences: 
firstly, since we are dealing with bosons, the Fock terms have the opposite 
sign compared to fermions; secondly, since a macroscopic condensate 
might appear, we should be cautious not to count  
twice terms between the same modes. 
We have avoided this difficulty by  
excluding carefully in the sum over the modes those leading to a 
double counting and which appear only in the integral terms.
Finally, another difference is that $n_{\vc{k}}$ is 
still a function depending on the time, 
generalizing in this way the RPA approach for systems in 
non equilibrium. 

For comparison, Zaremba et al. \cite{Zaremba} 
have derived a QKE in an approximation which would correspond 
to a different equation for $\rho_{\vc{k},\vc{q}}$. As we shall 
see below, their analysis is equivalent to removing the 
integral terms in (\ref{rho3}) and to  
approximating the quasi-particle energy by 
\begin{eqnarray}\label{ZNG}
\epsilon^{ZNG}_{\vc{k}}=\frac{\vc{k}^2}{2m}
+ \sum_{\vc{k'}}\frac{U_{\vc{0}}}{V}n_{\vc{k'}}+
\sum_{\vc{k'}\not=\vc{k}}\frac{U_{\vc{k}-\vc{k'}}}{V}n_{\vc{k'}}
\end{eqnarray}
We notice immediately that they have avoided a double 
counting in the exchange term. As a consequence, 
in the particular case of 
a contact potential and for $\vc{k_s}=0$, 
the energy difference that they obtained 
has a gap:
\begin{eqnarray}
\epsilon^{ZNG}_{\vc{k}}-\epsilon^{ZNG}_{\vc{0}}=
\frac{\vc{k}^2}{2m}+\g \frac{n_{\vc{0}}}{V}
\stackrel{\vc{k}\rightarrow 0}{\rightarrow}\g \frac{n_{\vc{0}}}{V}
\end{eqnarray}
Clearly, for $\vc{k} \rightarrow 0$ and if the 
mode $\vc{k_s}=0$ is macroscopically populated, 
this finite gap cannot be neglected in the 
scattering energy spectrum. 
In our approach, however, this gap does not appear 
anymore in $\epsilon^{HFA}_{\vc{k}}$.

The two operatorial equations (\ref{n}) and 
(\ref{rho3}) are the equations leading to the QKE.

\subsection{Collective and scattering excitations}

It is instructive to analyze the frequency 
spectrum solution of the eigenvalue problem 
given by Eq.(\ref{rho3}). In an electron liquid, 
the linear equation possesses two kinds of eigenvectors: 
(i) the scattering solutions involving the presence 
of only one particle and (ii) the collective 
solutions involving the presence of many particles.

For reasons we will explain below, let us concentrate on 
the problem of calculating the impulsion response or 
dielectric propagator 
${\cal U}_\vc{q}(\vc{k},\vc{k_1},t)$ to an initial 
particle having a momentum $\vc{k_1}$. 
We replace the operator $\rho_{\vc{k},\vc{q}}$ by this 
function and we substitute the interaction potential 
by its first order expression in (\ref{U0}). 
After simplifications, the integral 
equation for this function is: 
\begin{eqnarray}\label{U}
[i\frac{\partial}{\partial t}
-
\frac{\vc{k}.\vc{q}}{m}-\frac{\vc{q}^2}{2m}]
{\cal U}_\vc{q}(\vc{k},\vc{k_1},t)
\nonumber \\
=\g \big[\frac{(n_{\vc{k}}-n_{\vc{k+q}})}{V}
\sum_{\vc{k'}\not= \vc{k}}
{\cal U}_\vc{q}(\vc{k'},\vc{k_1},t)
\nonumber \\
+\frac{n_{\vc{k}}}{V}
\sum_{\vc{k'}\not=\vc{k}-\vc{q}}
{\cal U}_\vc{q}(\vc{k'},\vc{k_1},t)
-\frac{n_{\vc{k+q}}}{V}\sum_{\vc{k'}\not=\vc{k}+\vc{q}}
{\cal U}_\vc{q}(\vc{k'},\vc{k_1},t)\big]
\end{eqnarray} 
with the initial condition:
\begin{eqnarray}
{\cal U}_\vc{q}(\vc{k},\vc{k_1},t=0)=
\delta_{\vc{k},\vc{k_1}}
\end{eqnarray}

If $n_{\vc{k}}$ varies slowly in time and thus 
can be considered as constant during the evolution 
of the impulsion response, Eq.(\ref{rho3}) can be solved 
exactly in the thermodynamic limit. Similar equations have been solved 
in the field of plasma physics \cite{Ichimaru}. 
For this purpose,  we define 
the Laplace transform as: 
\begin{eqnarray}
{\cal U}_\vc{q}(\vc{k},\vc{k_1},\omega)
=\int_0^\infty \!\!\!dt\, e^{i(\omega+i0_+)t}{\cal U}_\vc{q}(\vc{k},\vc{k_1},t)
\end{eqnarray}
A $0_+$ has been added in order to ensure convergence of 
the integral. 
If we suppose that $\vc{k_s}$
is the wave-vector for the superfluid mode,  
then we can make the decomposition between a normal 
component and components involving the superfluid mode:
\begin{eqnarray}
{\cal U}_\vc{q}(\vc{k},\vc{k_1},\omega)=
{\tilde {\cal U}}_\vc{q}(\vc{k},\vc{k_1},\omega) 
+\delta_{\vc{k},\vc{k_s}}
{\cal U}_\vc{q}(\vc{k_s},\vc{k_1},\omega)
\nonumber \\
+\delta_{\vc{k},\vc{k_s}-\vc{q}}
{\cal U}_\vc{q}(\vc{k_s}-\vc{q},\vc{k_1},\omega)
\end{eqnarray}
Also, we distinguish the normal mode population 
$n'_{\vc{k}}=
(1-\delta_{\vc{k},\vc{k_s}})n_{\vc{k}}$ from the 
condensed mode.
Plugging this decomposition into (\ref{rho3}) 
and neglecting some $n'_{\vc{k}}$ by taking the 
thermodynamic limit,
we obtain for the superfluid modes:
\begin{eqnarray}\label{col1}
[\omega+i0_+ -
\frac{\vc{k_s}.\vc{q}}{m}-\frac{\vc{q}^2}{2m}]
{\cal U}_\vc{q}(\vc{k_s},\vc{k_1},\omega)
=i\delta_{\vc{k_s},\vc{k_1}} 
\nonumber \\
+\g \frac{n_{\vc{k_s}}}{V}
(\sum'_{\vc{k'}} 2\tilde{{\cal U}}_\vc{q}(\vc{k'},\vc{k_1},\omega)
+
{\cal U}_\vc{q}(\vc{k_s},\vc{k_1},\omega)
\nonumber\\
+
{\cal U}_\vc{q}(\vc{k_s}-\vc{q},\vc{k_1},\omega))
\end{eqnarray}
\begin{eqnarray}\label{col2}
[\omega+i0_+ -
\frac{\vc{k_s}.\vc{q}}{m}+\frac{\vc{q}^2}{2m}]
{\cal U}_\vc{q}(\vc{k_s}-\vc{q},\vc{k_1},\omega)
=i\delta_{\vc{k_s}-\vc{q},\vc{k_1}}
\nonumber \\
- \g \frac{n_{\vc{k_s}}}{V}
(\sum'_{\vc{k'}} 2\tilde{{\cal U}}_\vc{q}(\vc{k'},\vc{k_1},\omega)
+
{\cal U}_\vc{q}(\vc{k_s},\vc{k_1},\omega)
\nonumber \\
+
{\cal U}_\vc{q}(\vc{k_s}-\vc{q},\vc{k_1},\omega))
\end{eqnarray}
and for the normal component:
\begin{eqnarray}\label{col3}
[\omega+i0_+ -
\frac{\vc{k}.\vc{q}}{m}-\frac{\vc{q}^2}{2m}]
{\cal U}_\vc{q}(\vc{k_s}-\vc{q},\vc{k_1},\omega)
=i\delta_{\vc{k},\vc{k_1}}
\nonumber \\
+ \frac{8\pi a}{m} \frac{(n_{\vc{k}}-n_{\vc{k+q}})}{V}
(\sum'_{\vc{k'}} \tilde{{\cal U}}_\vc{q}(\vc{k'},\vc{k_1},\omega)
+
{\cal U}_\vc{q}(\vc{k_s},\vc{k_1},\omega)
\nonumber \\
+
{\cal U}_\vc{q}(\vc{k_s}-\vc{q},\vc{k_1},\omega))
\end{eqnarray}
The prime in the sum excludes terms involving 
the condensate mode. 
This close set of equations is solved in the Appendix A. 
For $\vc{k}\not= \vc{k_s}, \vc{k_s}-\vc{q}$, the scattering 
solutions are $\omega= \epsilon_{\vc{k}+\vc{q}}-\epsilon_{\vc{k}}$.
For excitations involving a superfluid mode 
$\vc{k}= \vc{k_s}, \vc{k_s}-\vc{q}$, the presence of 
interaction with the macroscopic condensate transforms 
the scattering solutions 
into  collective solutions of the 
discriminant equation:
\begin{eqnarray}\label{disprel}
\Delta(\vc{q},\omega)=
{\cal K}_n(\vc{q},\omega)[(\omega+i0_+ -
\frac{\vc{k_s}.\vc{q}}{m})^2 - {\epsilon^B_{\vc{q}}}^2]
\nonumber \\+
({\cal K}_n(\vc{q},\omega)-1)\frac{8\pi a n_{\vc{k_s}}}{mV}
\frac{\vc{q}^2}{m}=0
\end{eqnarray}
where 
\begin{eqnarray}
\epsilon^B_{\vc{q}}=
\sqrt{c^2 \vc{q}^2 +
(\frac{\vc{q}^2}{2m})^2}
\end{eqnarray}
is the Bogoliubov excitation energy,
\begin{eqnarray}
c=\sqrt{\frac{4\pi a n_{\vc{k_s}}}{m^2 V}} 
\end{eqnarray}
is the sound velocity
and 
\begin{eqnarray}\label{kn}
{\cal K}_n(\vc{q},\omega)=
1- \frac{8\pi a }{mV}\sum_{\vc{k}}
\frac{n'_{\vc{k}}-n'_{\vc{k+q}}}{
\omega+i0_+ -
\frac{\vc{k}.\vc{q}}{m}-\frac{\vc{q}^2}{2m}}
\end{eqnarray}
is the dynamic dielectric function of the normal 
fluid. 
Eq.(\ref{disprel}) allows to find a 
dispersion relation for any 
function $n_{\vc{k}}$ possibly in non thermodynamic 
equilibrium. 
In this sense,
it generalizes the dispersion relation 
that is obtained equivalently from 
the density fluctuations response formalism for 
a gas at equilibrium \cite{SK,PN,GriffinB,Minguzzi,Giorgini}. 
Note that in \cite{Giorgini} 
the non number conserving 
Beliaev formalism has been used to derive the dispersion 
relation up to the next order beyond the Bogoliubov theory.
The solution 
can be put in the complex form: 
$\omega=\omega_{\vc{q}}-i\gamma_{\vc{q}}$ 
where $\gamma_{\vc{q}}$ corresponds to the Landau damping.
For the particular 
case of a weakly depleted Bose gas, we can solve 
analytically Eq.(\ref{disprel}). In that case, we 
can approximate \cite{SK}:
\begin{eqnarray} \label{Knapp}
{\cal K}_n(\vc{q},\omega) \simeq 
1+i{\rm Im} {\cal K}_n(\vc{q},\omega)
\end{eqnarray}
Eq.(\ref{disprel}) is obtained considering 
in a first approximation that the imaginary 
term can be neglected. We find that the real 
part corresponds to the Bogoliubov 
spectrum:
\begin{eqnarray}\label{reo}
\omega_{\vc{q}} \simeq \frac{\vc{k_s}.\vc{q}}{m} 
\pm \epsilon^B_{\vc{q}}
\end{eqnarray}
The imaginary part corresponds to the 
Landau damping and can be calculated 
perturbatively assuming 
$|\gamma_{\vc{q}}| \ll \omega_{\vc{q}}$ which is the 
case for a weakly depleted condensate. We find, up to the 
first order, 
\begin{eqnarray}\label{imo}
\gamma_{\vc{q}} \simeq 
{\rm Im}{\cal K}_n(\vc{q},\omega_{\vc{q}})
\frac{\frac{4\pi a n_{\vc{k_s}}}{mV}
\frac{\vc{q}^2}{m}}{\omega_{\vc{q}}-\frac{\vc{k_s}.\vc{q}}{m}}
\end{eqnarray}
By analogy with plasma physics, we say that a Bose gas 
is stable provided that $\gamma_{\vc{q}} \geq 0$ and unstable 
otherwise. In a stable condensate the collective oscillations 
are damped, while in a unstable condensate they are amplified 
exponentially.  

For the case of thermal equilibrium (\ref{neq})  
with $\vc{v_n}=0$, $\mu=0$ and a temperature close to zero, 
${\rm Im}{\cal K}_n(\vc{q},\omega_{\vc{q}})$ 
is calculated in appendix B and is a positive function for 
$\omega_{\vc{q}} \geq 0$ and negative otherwise. Thus, from 
(\ref{imo}) the stability condition could be written as:
\begin{eqnarray}\label{stab}
\epsilon^B_{\vc{q}}-\frac{\vc{k_s}.\vc{q}}{m} \geq 0
\end{eqnarray}
We can check that this inequality is fulfilled for all values of 
the momentum at the condition that 
$|\vc{k_s}/m| \leq c$. In other words, the condensate is 
stable if its velocity relative to the normal fluid is 
much less than the sound velocity. Anticipating the next 
subsections, this condition corresponds to the Landau 
criterion for superfluidity for a weakly depleted Bose gas. 
More generally, the occurrence of instability depends on 
the form of $n_\vc{k}$ which influences the spectrum 
obtained from (\ref{disprel}). In the case of a plasma, 
this problem has been studied long time ago 
\cite{instB,FR}.
  
Finally, in the absence of a macroscopic condensate 
i.e. $n_\vc{k_s}/V \rightarrow 0$, 
we recover 
the two scattering solutions for the superfluid:
\begin{eqnarray}
\omega
\stackrel{n_{\vc{k_s}}\rightarrow 0}{=}
\frac{\vc{k_s}.\vc{q}}{m}\pm\frac{\vc{q}^2}{2m}
\end{eqnarray}
and the collective solution is given by 
\begin{eqnarray}
{\cal K}_n(\vc{q},\omega)=0
\end{eqnarray}
According to \cite{SK}, at equilibrium,
this collective solution contains only an imaginary 
part and thus no collective oscillation can be 
observed in the 
system. Therefore, any collective oscillation results 
specifically from the condensation which transforms 
the scattering solution of the condensed mode 
into a collective solution. 

\subsection{Derivation of the kinetic equation}

From the results of the previous section, 
we are now ready to derive a generalized Boltzmann like 
equation for the Bose condensed gas. We  
define the spatial correlation function as 
the average $\langle 
\rho_{\vc{k'},-\vc{q}}\rho_{\vc{k},\vc{q}} \rangle$. 
Using Eq.(\ref{rho3}), we derive the 
following equation for the correlation function:
\begin{eqnarray}\label{K1}
[i\frac{\partial}{\partial t}
+
\frac{(\vc{k'}-\vc{k}).\vc{q}}{m}-\frac{\vc{q}^2}{m}]
\langle \rho_{\vc{k'},-\vc{q}} \rho_{\vc{k},\vc{q}}
\rangle
\nonumber \\
=\frac{4\pi a}{mV} \sum_{\vc{k''}}\big[n_{\vc{k}}
(2-\delta_{\vc{k},\vc{k''}}-\delta_{\vc{k}-\vc{q},\vc{k''}})
\nonumber \\
-n_{\vc{k+q}}
(2-\delta_{\vc{k},\vc{k''}}-\delta_{\vc{k}+\vc{q},\vc{k''}})
\big]
\langle \rho_{\vc{k'},-\vc{q}} \rho_{\vc{k''},\vc{q}}
\rangle
\nonumber \\
+\big[
n_{\vc{k'}}
(2-\delta_{\vc{k'},\vc{k''}}-\delta_{\vc{k'}+\vc{q},\vc{k''}})
\nonumber \\
-n_{\vc{k'-q}}
(2-\delta_{\vc{k'},\vc{k''}}-\delta_{\vc{k'}-\vc{q},\vc{k''}})
\big] \langle \rho_{\vc{k''},-\vc{q}}\rho_{\vc{k},\vc{q}}
\rangle
\end{eqnarray}

This function 
can be decomposed as a non interacting part and an 
interacting part \cite{Balescu}:
\begin{eqnarray}\label{g}
\langle \rho_{\vc{k'},-\vc{q}}
\rho_{\vc{k},\vc{q}} \rangle=
(n_{\vc{k}}+1)n_{\vc{k'}}
\delta_{\vc{k'}-\vc{k},\vc{q}}
\nonumber\\
+n_{\vc{k}}n_{\vc{k'}}
\delta_{\vc{q},0}(1-\delta_{\vc{k},\vc{k'}})
-n_{\vc{k}}\delta_{\vc{k},\vc{k'}}\delta_{\vc{q},0}
+g_\vc{q}(\vc{k},\vc{k'})
\end{eqnarray}
where $g_\vc{q}(\vc{k},\vc{k'})$ 
represents the correlation function due 
to the interactions. For $\vc{q} \not= 0$ or $\vc{k} \not= \vc{k'}$,
the non interacting part follows the Wick's decomposition and 
since the system is homogeneous $\langle c^\dagger_\vc{k} 
c_\vc{k'} \rangle = \delta_{\vc{k},\vc{k'}} n_\vc{k}$. 
For $\vc{q}= 0$ or $\vc{k}=\vc{k'}$, however, it  
reduces to the quadratic average 
population that we choose to be $\langle {\hat n}_\vc{k}^2 \rangle =
n_\vc{k}^2$. 

Would we have used the Wick's decomposition in that 
case then $\langle {\hat n}_\vc{k}^2 \rangle =
2 n_\vc{k}^2$ which corresponds to non zero particle number 
fluctuations $\langle \delta^2 {\hat n}_\vc{k} \rangle =
n_\vc{k}^2$ and the total particle number would display fluctuations 
as well. Indeed, since 
$\langle {\hat n}_\vc{k} {\hat n}_\vc{k'}\rangle= n_\vc{k}n_\vc{k'}$ 
for $\vc{k}\not=\vc{k'}$, 
we calculate $\langle \delta^2 {\hat N} \rangle=
\sum_{\vc{k}} n_\vc{k}^2$. In the presence of condensation, 
these fluctuations are huge i.e. of the same 
order of the average value \cite{4eme}. 
Consequently, in an isolated system where the 
total particle number is conserved, this unphysical 
situation must be excluded. 

Taking the average over each side of
Eq.(\ref{n}), a comparison with (\ref{g}) allows to deduce:
\begin{eqnarray}\label{n2}
i\frac{\partial}{\partial t}n_{\vc{k}}
=\sum_{\vc{q},\vc{k'}}\frac{2\pi a}{mV}
\big(g_\vc{q}(\vc{k},\vc{k'})-
g_\vc{q}(\vc{k}-\vc{q},\vc{k'})
\nonumber \\
+g_\vc{q}(\vc{k'},\vc{k})-
g_\vc{q}(\vc{k'},\vc{k}+\vc{q})\big)
\end{eqnarray}
On the other hand,
inserting this definition into Eq.(\ref{K1}) and using 
(\ref{n2}), 
we obtain 
\begin{eqnarray}\label{K2}
[i\frac{\partial}{\partial t}
+
\frac{(\vc{k'}-\vc{k}).\vc{q}}{m}-\frac{\vc{q}^2}{m}]
g_\vc{q}(\vc{k},\vc{k'})
\nonumber \\
=Q_\vc{q}(\vc{k},\vc{k'})+
\frac{4\pi a}{mV} \sum_{\vc{k''}}
 \big[n_{\vc{k}}
(2-\delta_{\vc{k},\vc{k''}}-\delta_{\vc{k}-\vc{q},\vc{k''}})
\nonumber \\
-n_{\vc{k+q}}
(2-\delta_{\vc{k},\vc{k''}}-\delta_{\vc{k}+\vc{q},\vc{k''}})
\big]g_\vc{q}(\vc{k''},\vc{k'})
\nonumber \\
+\big[n_{\vc{k'}}
(2-\delta_{\vc{k'},\vc{k''}}-\delta_{\vc{k'}+\vc{q},\vc{k''}})
\nonumber \\
-n_{\vc{k'-q}}
(2-\delta_{\vc{k'},\vc{k''}}-\delta_{\vc{k'}-\vc{q},\vc{k''}})
\big]g_\vc{q}(\vc{k},\vc{k''})
\end{eqnarray}
where 
\begin{eqnarray}\label{Q}
Q_\vc{q}(\vc{k},\vc{k'})=
\frac{8\pi a }{mV}
\big[(n_{\vc{k}}-n_{\vc{k+q}})(n_{\vc{k'-q}}+1)n_{\vc{k'}}
\nonumber \\
+(n_{\vc{k'}}-n_{\vc{k'-q}})(n_{\vc{k}}+1)n_{\vc{k+q}}\big]
\nonumber \\
-\frac{4\pi a n_{\vc{k_s}}^2}{mV}(
\delta_{\vc{k},\vc{k_s}}
\delta_{\vc{k'},\vc{k_s}}
-\delta_{\vc{k},\vc{k_s}-\vc{q}}
\delta_{\vc{k'},\vc{k_s}+\vc{q}})
\nonumber \\
\left[(1+n_{\vc{k_s}})n_{\vc{k_s}+\vc{q}}+(1+n_{\vc{k_s}-\vc{q}})n_{\vc{k_s}}
\right]
\end{eqnarray}
is the inhomogeneous term.
To get (\ref{Q}), we eliminate terms involving delta 
functions which will give negligible contribution to the QKE in the 
thermodynamic limit. 
Both  Eq.(\ref{n2}) and Eq.(\ref{K2}) form a close set in which 
$g_\vc{q}(\vc{k},\vc{k'})$ must be eliminated 
in order to get a kinetic equation for $n_{\vc{k}}$. 
This elimination is done following an analog procedure 
to that of Ichimaru \cite{Ichimaru}. 
We assume that the  
correlations due to the interactions are 
non-existent for $t \rightarrow -\infty$. This requirement is usual 
in  kinetic theory and allows to provide the 
following initial condition
$g_\vc{q}(\vc{k},\vc{k'})|_{t \rightarrow -\infty} =0$.
Also, we assume that $n_{\vc{k}}$ evolves on a much 
more long time scale than the duration of a collision 
$g_\vc{q}(\vc{k},\vc{k'})$ and 
so is considered as constant in solving Eq.(\ref{K2}). 
This approximation amounts to claiming that the binary
collision process is instantaneous in comparison
with the time associated to the relaxation of the system.
Inspired by the previous subsections and by \cite{Ichimaru},
we can check that the solution, 
satisfying both Eq.(\ref{K2}) and the initial condition,
is expressed in terms of the dielectric propagator as:
\begin{eqnarray}\label{K3}
g_\vc{q}(\vc{k},\vc{k'},t)=
-i\int_0^\infty \!\!\!\!\!dt' \sum_{\vc{k_1},\vc{k'_1}}
\nonumber \\
{\cal U}_{\vc{q}}(\vc{k},\vc{k_1},t')
{\cal U}_{-\vc{q}}(\vc{k'},\vc{k'_1},t')
Q_\vc{q}(\vc{k_1},\vc{k'_1},t-t')
\end{eqnarray}
We have inserted the explicit time dependence in the functions.
If the creation of such correlations is 
much faster in comparison with the relaxation time for $n_{\vc{k}}$, 
$Q_\vc{q}(\vc{k_1},\vc{k'_1},t-t') \simeq 
Q_\vc{q}(\vc{k_1},\vc{k'_1},t)$ and we obtain a 
Markovian equation. 
We re express the dielectric 
propagator in Fourier transform according to, 
\begin{eqnarray}
{\cal U}_\vc{q}(\vc{k},\vc{k_1},t)
=\frac{1}{2\pi}
\int_{\cal C} d\omega e^{-i\omega t}
{\cal U}_\vc{q}(\vc{k},\vc{k_1},\omega)
\end{eqnarray}
where the contour ${\cal C}$ extends from $-\infty$ to 
$+\infty$ along a path in the upper half of the $\omega$ plane 
in such a way that all the singularities lie below it. 
The substitution allows to carry out successively integrations 
over $t'$ and $\omega'$ by closing the contour in the 
upper half plane  in order to get:
\begin{eqnarray}\label{K4}
g_\vc{q}(\vc{k},\vc{k'})=
\int_{-\infty}^\infty \!\!\!\frac{d\omega}{2\pi i}\sum_{\vc{k_1},\vc{k'_1}}
\nonumber \\
{\cal U}_{\vc{q}}(\vc{k},\vc{k_1},\omega)
{\cal U}_{-\vc{q}}(\vc{k'},\vc{k'_1},-\omega)
Q_\vc{q}(\vc{k_1},\vc{k'_1},t)
\end{eqnarray}
Calculations in appendix C allow to find an explicit 
expression for $g_\vc{q}(\vc{k},\vc{k'})$ in terms of 
the one particle distribution function, provided  
the Landau damping factor is always positive. 
The substitution into 
(\ref{n2}) allows finally to get the GRPA kinetic 
equation for a stable Bose gas:
\begin{eqnarray}\label{K5}
\frac{\partial}{\partial t}n_{\vc{k}}=
{\cal C}^T_{\vc{k}}[n_{\vc{k'}};\vc{k_s}]=
{\cal C}_{\vc{k}}[n_{\vc{k'}};\vc{k_s}]+
{\tilde {\cal C}}_{\vc{k}}[n_{\vc{k'}};\vc{k_s}]
\end{eqnarray}
where we define the collision terms as a functional of $n_{\vc{k'}}$
and a function of $\vc{k_s}$. The first term describes the 
collision rate between particles 
of the normal fluid:
\begin{widetext}
\begin{eqnarray}\label{K6}
{\cal C}_{\vc{k}}[n_{\vc{k'}};\vc{k_s}]
=\sum_{\vc{q},\vc{k'}} \left|
\frac{\frac{8\pi a}{mV}}
{{\cal K}(\vc{q},\epsilon_{\vc{k}+\vc{q}}-\epsilon_{\vc{k}})}
\right|^2 \!\!
(1-\delta_{\vc{k},\vc{k_s}}-\delta_{\vc{k}+\vc{q},\vc{k_s}}-
\delta_{\vc{k'},\vc{k_s}}-\delta_{\vc{k'}-\vc{q},\vc{k_s}})
\pi \delta(\epsilon_{\vc{k}+\vc{q}}+\epsilon_{\vc{k'}-\vc{q}}-
\epsilon_{\vc{k}}-\epsilon_{\vc{k'}})
\nonumber \\
\big[n_{\vc{k}+\vc{q}}n_{\vc{k'}-\vc{q}}
(n_{\vc{k}}+1)(n_{\vc{k'}}+1)-
n_{\vc{k}}n_{\vc{k'}}(n_{\vc{k}+\vc{q}}+1)(n_{\vc{k'}-\vc{q}}+1)\big]
\end{eqnarray}
The second term describes the collision rate between condensed
and non condensed particles, the condensed particle is
either the input or the output state in the scattering
process:
\begin{eqnarray}\label{K7}
{\tilde{\cal C}}_{\vc{k}}[n_{\vc{k'}};\vc{k_s}]
=&\displaystyle \sum_{\vc{q},\vc{k'}}
\frac{\left(\frac{8\pi a}{mV}\right)^2}
{\left|{\cal K}^*(\vc{q},\epsilon_{\vc{k}+\vc{q}}-\epsilon_{\vc{k}})
{\tilde{\cal K}}(\vc{q},\epsilon_{\vc{k}+\vc{q}}-\epsilon_{\vc{k}})\right|}
(\delta_{\vc{k},\vc{k_s}}+\delta_{\vc{k}+\vc{q},\vc{k_s}}+
\delta_{\vc{k'},\vc{k_s}}+\delta_{\vc{k'}-\vc{q},\vc{k_s}})
\nonumber \\
&\pi \delta(\epsilon_{\vc{k}+\vc{q}}+\epsilon_{\vc{k'}-\vc{q}}-
\epsilon_{\vc{k}}-\epsilon_{\vc{k'}})
\big[n_{\vc{k}+\vc{q}}n_{\vc{k'}-\vc{q}}
(n_{\vc{k}}+1)(n_{\vc{k'}}+1)-
n_{\vc{k}}n_{\vc{k'}}(n_{\vc{k}+\vc{q}}+1)(n_{\vc{k'}-\vc{q}}+1)\big]
\end{eqnarray}
\end{widetext}
In the first expression (\ref{K6}), the contact potential has 
been replaced by an effective one depending 
on the transfer particle energy 
$\omega=\epsilon_{\vc{k}+\vc{q}}-\epsilon_{\vc{k}}$:
\begin{eqnarray}\label{ueff}
U^{eff}_{\vc{q}}(\omega)
=\frac{\frac{8\pi a}{mV}}
{{\cal K}(\vc{q},\omega)}
\end{eqnarray}
The correcting term is the dynamic dielectric constant:
\begin{eqnarray}\label{kal}
&&{\cal K}(\vc{q},\omega)
\nonumber \\
&=&
\frac{\frac{8\pi a n_{\vc{k_s}}}{mV}
\frac{\vc{q}^2}{m}}
{(\omega+i0_+ -
\frac{\vc{k_s}.\vc{q}}{m})^2-(\frac{\vc{q}^2}{2m})^2+
\frac{4\pi a n_{\vc{k_s}}}{mV}
\frac{\vc{q}^2}{m}}+{\cal K}_n(\vc{q},\omega)
\nonumber \\
&=&
\frac{\Delta(\vc{q},\omega)}
{(\omega+i0_+ -\frac{\vc{k_s}.\vc{q}}{m})^2-(\frac{\vc{q}^2}{2m})^2+
\frac{4\pi a n_{\vc{k_s}}}{mV}
\frac{\vc{q}^2}{m}}
\end{eqnarray}
In the second expression (\ref{K7}), 
we define another dynamical dielectric function:
\begin{eqnarray}\label{kalc}
{\cal \tilde{K}}(\vc{q},\omega)
=\frac{\Delta(\vc{q},\omega)}
{(\omega+i0_+ -\frac{\vc{k_s}.\vc{q}}{m})^2-(\frac{\vc{q}^2}{2m})^2}
\end{eqnarray}
This result generalizes (\ref{heur}) in the case where 
$n'_{\vc{k}} \not= 0$.
Plugging this expression into Eq.(\ref{K7}), the energy 
conservation imposes the two choices 
$\omega=\frac{\vc{k_s}.\vc{q}}{m} \pm \frac{\vc{q}^2}{2m}$
when a particle of the condensate participates to the collision 
process. But, 
with such an energy transfer and for a macroscopic population 
of the condensate, this dynamic dielectric function gets 
infinite: 
\begin{eqnarray}\label{kalc2}
{\cal \tilde{K}}(\vc{q},\omega)
|_{\omega=\frac{\vc{k_s}.\vc{q}}{m} \pm \frac{\vc{q}^2}{2m}}
\rightarrow \infty
\end{eqnarray}
leading to ${\tilde{\cal C}}_{\vc{k}}[n_{\vc{k'}};\vc{k_s}]
\rightarrow 0$.
In this situation, surprisingly, no collision involving the condensate particle
occurs. As a consequence, in a uniform Bose gas, no transfer
of particle is possible between the condensate and the
normal component. Indeed, due to the
absence of Fock interaction term in the GRPA, the effective
potential has a different expression when a particle
of the condensate is involved in a scattering.
As said in section 2, a potential induced by the 
condensate compensates 
exactly the potential created by the excited particle susceptible 
to scatter. 
As a consequence, the dielectric function suppresses 
the effectiveness of the  
potential which thus becomes  
invisible to the condensate and totally shelters it from collision.
Furthermore, let us note that this 
shielding remains unchanged for any momentum $\vc{k_s}$ 
which preserves the stability of the condensate. If 
the Landau damping factor $\gamma_{\vc{q}}$ becomes negative, 
then an instability occurs and therefore Eq.(\ref{K6}) and Eq.(\ref{K7}) 
are not longer valid, as well as all considerations concerning 
collision blockade. 
In that case, a more elaborated 
derivation must be carried out that 
could allow particle exchange with the condensate.
The expressions (\ref{B2}) and 
(\ref{B3}) must be recalculated taking into account the 
instability \cite{instB,FR}.

In this way, the GRPA kinetic equation provides a 
different understanding of
the superfluidity phenomenon as it describes the motion of
two independent fluids that cannot transfer particle through
collisions. Since $\frac{\partial}{\partial t}n_{\vc{k_s}}=0$,
$n_\vc{k_s}$ is really a constant of motion independent of the 
dynamics of the normal fluid.

The collision blockade is the consequence of a higher 
order expansion in the interaction parameter. Since 
${\cal \tilde{K}}(\vc{q},\omega)|_{a \rightarrow 0}=1$, 
up to the second order in the interaction,  
we recover the UUQKE
allowing collisions with the condensate.
Only an infinite re-summation of appropriate higher order 
contributions through GRPA generates collectively a perfect collision 
blockade. Also, in the limit where the condensate is not 
macroscopically  populated $n_{\vc{k_s}}/V \rightarrow 0$, 
then ${\cal \tilde{K}}(\vc{q},\omega) \rightarrow 
{\cal K}_n(\vc{q},\omega)$ and 
${\cal K}(\vc{q},\omega)\rightarrow
{\cal K}_n(\vc{q},\omega)$ and collisions with 
particle of the mode $\vc{k_s}$ become again possible.
The Eq.(\ref{K5}) is a different 
version of the kinetic equations formulated by
Zaremba et al. \cite{Zaremba} in which  
collisions are possible with the condensate particle. 
Indeed, in their approach, the authors have done an 
approximation which is equivalent to 
replacing the energy $\epsilon_{\vc{k}}$ by (\ref{ZNG}) 
and to  
suppressing the 
integral terms of the right hand side in (\ref{rho3}). 
As a consequence, they obtained a QKE 
in which
the dielectric function
does not appear and 
with the 
feature of having a gap in the transition of a 
particle to the condensate.

Let us notice that the same phenomenon occurs for the 
simple RPA where the exchange terms have been omitted. 
In the SRPA the calculation is similar to those
leading to the quantum kinetic equation for a plasma.
For details of such derivation, see Balescu \cite{QBalescu}.
The only difference is that
the coulombian potential becomes
a contact potential and that the fermion now becomes
a boson, meaning that any factor
$1-n_{\vc{k}}$ now becomes $1+n_{\vc{k}}$ in the
collision term.
In that case, the dielectric function has the 
same form for both the condensed and the normal modes \cite{SK}:
\begin{eqnarray}\label{ksrpa}
{\cal K}^{SRPA}(\vc{q},\omega)
&=&1- \frac{4\pi a }{mV}\sum_{\vc{k}}
\frac{n_{\vc{k}}-n_{\vc{k+q}}}{
\omega+i0_+ -
\frac{\vc{k}.\vc{q}}{m}-\frac{\vc{q}^2}{2m}}
\nonumber \\
&=&
\frac{\frac{4\pi a n_{\vc{k_s}}}{mV}
\frac{\vc{q}^2}{m}}
{(\omega+i0_+ -
\frac{\vc{k_s}.\vc{q}}{m})^2-(\frac{\vc{q}^2}{2m})^2}
\nonumber \\
&&+{\cal K}^{SRPA}_n(\vc{q},\omega)
\end{eqnarray} 
The normal dielectric component 
${\cal K}^{SRPA}_n(\vc{q},\omega)$ is of the 
same form as Eq.(\ref{kn}), except that the exchange 
interaction term 
has been removed.
Similarly, the particles of the condensate are protected from 
any collision with the surrounding. Nevertheless, in the SRPA, 
collisions involving 
low level excited states are strongly prevented because the energy 
transfer $\omega$ is close to the one corresponding to the 
condensate. On the contrary, in the GRPA, 
a supplementary term coming from the exchange effect 
$\frac{4\pi a n_{\vc{k_s}}}{mV}$ ensures a reasonable value 
for the effective potential guaranteeing an efficient collision rate, 
even with particle in the energy levels close to the condensate level.

\subsection{Properties of the RPA collision term}

The collision term exhibits a number of remarkable 
properties analog to those encountered by other kinetic 
equations \cite{Ichimaru}. 
These allow to establish the particle number, momentum and
energy conservation laws as well as the Boltzmann H-theorem.
Since the condensed particles 
do not participate to the collision process,
these concerns only the excited particles of the normal fluid. 
Three of them 
can be stated as:
\begin{eqnarray}
\sum'_{\vc{k}}{\cal C}_{\vc{k}}[n_{\vc{k'}};\vc{k_s}] =0
\\
\sum'_{\vc{k}} \vc{k} {\cal C}_{\vc{k}}[n_{\vc{k'}};\vc{k_s}]=0
\\
\sum'_{\vc{k}} \epsilon_{\vc{k}}{\cal C}_{\vc{k}}
[n_{\vc{k'}};\vc{k_s}]=0
\end{eqnarray}
These three integral relations are checked easily by dividing 
the collision term in four 
equal parts.  After carrying out the successive changes of 
the integration variables $\vc{k} \leftrightarrow \vc{k'}, 
\, \vc{q} \leftrightarrow -\vc{q}$ on the second term,
$\vc{k} \leftrightarrow \vc{k'}-\vc{q}, \, 
\vc{k'} \leftrightarrow \vc{k}+ \vc{q}$ on the third term and 
a successive combination of these two variable changes in the 
fourth term, these four terms cancel 
each other if we use the relation 
${\cal{K}}(\vc{q},\omega)={\cal{K}}^\star(-\vc{q},-\omega)$.
As a consequence, for a uniform Bose gas, the total 
particle number, the total momentum and the total kinetic energy, 
associated 
with the normal component are independent of the time i.e.
$(d/dt)\sum_{\vc{k}} n'_{\vc{k}}=0$, 
$(d/dt)\sum_{\vc{k}} \vc{k}n'_{\vc{k}}=0$ and 
$(d/dt)\sum_{\vc{k}} \epsilon_{\vc{k}} n'_{\vc{k}}=0$.
These properties together with the conservation of 
$n_\vc{k_s}$ imply that the total energy in the 
Hartree-Fock approximation is also conserved \cite{Zaremba} 
i.e.:
\begin{eqnarray}
\frac{d}{dt}
[\sum_{\vc{k}} \epsilon_{\vc{k}} n'_{\vc{k}} +
\frac{4\pi a}{mV}
(N^2 - \frac{1}{2} n^2_\vc{k_s})]=0 
\end{eqnarray}  

Another crucial property is the Boltzmann H-theorem.
The entropy, due to thermal excitations, has the expression:
\begin{eqnarray}\label{S}
S=\sum'_{\vc{k}}[(1+n'_{\vc{k}})\ln (1+n'_{\vc{k}})-
n'_{\vc{k}}\ln n'_{\vc{k}}]
\end{eqnarray}
The time evolution of the entropy is always positive. A 
similar derivation using the change of variables allows to 
calculate a positive production of entropy:
\begin{widetext}
\begin{eqnarray}\label{H}
\frac{dS}{dt}
=\!\! \sum_{\vc{q},\vc{k},\vc{k'}}
\left|
\frac{\frac{8\pi a}{mV}}
{{\cal K}(\vc{q},\epsilon_{\vc{k}+\vc{q}}-\epsilon_{\vc{k}})}
\right|^2 \!\!
\frac{\pi}{4}
\delta(\epsilon_{\vc{k}+\vc{q}}+\epsilon_{\vc{k'}-\vc{q}}-
\epsilon_{\vc{k}}-\epsilon_{\vc{k'}})
\ln\left[\frac{n'_{\vc{k}+\vc{q}}n'_{\vc{k'}-\vc{q}}
(n'_{\vc{k}}+1)(n'_{\vc{k'}}+1)}{
n'_{\vc{k}}n'_{\vc{k'}}(n'_{\vc{k}+\vc{q}}+1)(n'_{\vc{k'}-\vc{q}}+1)}\right]
\nonumber \\
\left[n'_{\vc{k}+\vc{q}}n'_{\vc{k'}-\vc{q}}
(n'_{\vc{k}}+1)(n'_{\vc{k'}}+1)-
n'_{\vc{k}}n'_{\vc{k'}}(n'_{\vc{k}+\vc{q}}+1)(n'_{\vc{k'}-\vc{q}}+1)\right]
\geq 0
\end{eqnarray}
\end{widetext}
The positivity is a consequence of the mathematical relation 
$\ln(x/y)(x-y)\geq 0$. The equality is achieved for $x=y$. 
From Eq.(\ref{H}) we deduce that the entropy always increases 
until the system reaches a stationary equilibrium distribution. 
This occurs when the 
production of entropy becomes zero. In that situation, 
\begin{eqnarray}
\frac{n'_{\vc{k}+\vc{q}}n'_{\vc{k'}-\vc{q}}}
{(n'_{\vc{k}+\vc{q}}+1)(n'_{\vc{k'}-\vc{q}}+1)}
=\frac{n'_{\vc{k}}n'_{\vc{k'}}}{(n'_{\vc{k}}+1)(n'_{\vc{k'}}+1)}
\end{eqnarray}
This relation holds only for the Bose-Einstein distribution 
function \cite{Balescu}:
\begin{eqnarray}\label{neq}
n'_{\vc{k}}=n^{eq}_{\vc{k}}=
\frac{1}{\exp{[\beta(\epsilon_\vc{k}-\vc{k}.\vc{v_n}-\mu)]}-1}
\end{eqnarray}
where $\vc{v_n}$ is the average velocity of the normal component. 
In this way, the inverse temperature $\beta$ and the chemical 
potential $\mu$ are defined as the free parameters of the equilibrium 
solution. 
Although similar, the properties of the thermodynamic 
equilibrium predicted by the H-theorem in the RPA are really different 
from the one predicted by the calculation of 
the ensemble partition functions of an ideal Bose gas \cite{Huang}.
The superfluid and normal components can move 
relatively with two different velocities $\vc{v_s}$ and 
$\vc{v_n}$.
The relative difference 
$\vc{v_s}-\vc{v_n}$ and $\mu$ are subject to the only constraint of 
stability which gives a limitation. 
As said in the previous subsection, for $\vc{v_n}=0$, 
and 
the temperature close to zero, and using (\ref{kn4}) in 
(\ref{imo}), this constraint 
is realized provided that $\mu=0$ and that 
the relative velocity does not 
exceed the sound velocity 
(see Eq.(\ref{stab})). Otherwise, the condensate becomes 
instable. Thus, we recover the Landau 
criterion for the weakly interacting Bose gas. 
Note that this result differs from the ZNG approach since 
it corresponds to an equilibrium fugacity of the normal fluid 
equal to one 
(see Eq.(44) in \cite{Zaremba}). 

On the contrary, the statistical equilibrium ensemble formalism  
for an ideal Bose gas imposes a zero relative velocity 
and a chemical potential close to zero when condensation 
occurs \cite{Huang}. 
This contradiction might be explained if we remind that 
in this formalism we postulate an equal {\it a priori}
probability of any possible configuration of the gas \cite{Huang, Balescu}. 
This basic assumption of equilibrium statistical 
mechanics originates from the observation that, 
over a long 
time range, the collision process will mix statistically 
these configurations. But, since in the GRPA kinetic equation the 
collision is blocked, the equal {\it a priori}
probability might not work as far as the condensate mode is concerned.
This observation suggests that the condensed  
particle population for an homogeneous systems 
is not distributed anymore but has a well 
defined value. 

\subsection{Condensate formation}

The kinetic equation (\ref{K5}) does not 
explain the condensate formation through, for 
example, evaporative cooling \cite{Keterlee2,BZS,Gardiner2}. 
If we start indeed from an initially 
condensed gas with a Bose-Einstein
momentum distribution whose the tail has been cut, then 
according to the RPA model, the irreversible
evolution process towards equilibrium 
will undergo a strange behavior. 
Instead of populating the lowest energy level which is 
forbidden, the excited particles will be transferred to the 
lowest excited levels in which a macroscopic population 
can eventually appear. In that situation, we are led 
to a fragmentation of the condensate into at least two 
energy levels. But such two macroscopic states contradict 
the assumption made at the beginning that only one macroscopic 
state does exist. A more elaborated model can take into account 
more than one level macroscopically populated. Although this 
more general description must not be excluded, 
for energetic reason, this phenomenon is not likely 
to happen \cite{GSS}. A fragmented condensate has a 
much higher potential energy than a non-fragmented one 
due to the presence of a Fock interaction energy between 
two fragmented parts. 
To explain the condensate formation and if we exclude 
fragmentation, we must assume either that the Bose gas 
must be instable due to a far non-equilibrium situation 
or that it must be inhomogeneous with regions of strong depletion. 
Indeed, in those regions - like the edge of the gas - the condensate 
population is not macroscopic anymore and so collisions between 
the two fluids may happen.

\section{Extension to a weakly inhomogeneous Bose gas}

\subsection{Inhomogeneous equations}

The extension of the previous equations to 
inhomogeneity 
is important in order to understand how 
the evolution of the condensate and  
the thermal excitations are coupled through 
mean field forces. 
In what follows, we assume that the gas 
is weakly inhomogeneous i.e. most of  
the quasi-particles collide inside a sufficiently small 
volume that can be considered 
as homogeneous. 
To quantify the level of acceptable 
inhomogeneity, we divide the volume $V$ of the gas 
into small cubic volume $\Omega$. The edge $l$ of each  
volume $\Omega=l^3$  is adjusted in such a way to be 
infinitesimal so that the gas is  
homogeneous inside it, but big enough to
contain a large amount of particle 
whose dynamic obeys still locally the homogeneous equation. 
In the literature \cite{Balescu}, 
$l$ is referred as the hydrodynamic  scale 
and is estimated from the formula 
$l(\vc{r},t)= \rho_e(\vc{r},t)/\nabla_{\vc{r}} \rho_e(\vc{r},t)$ 
where $\rho_e(\vc{r},t)$ is the normal fluid density  
and $\vc{r}$ 
is the position in space of the small 
volume. This length scale must be much greater than 
the mean free path 
which can be estimated more less as $1/(\rho_e\sigma)$
where  
$\sigma=8\pi a^2$ is the total cross section \cite{Balescu}.

In this infinitesimal volume, we define 
the local condensate wave function
\begin{eqnarray}
\Psi(\vc{r},t)=\sqrt{n_c(\vc{r},t)}
e^{i\theta (\vc{r},t)}
\end{eqnarray}
which corresponds to the eigenfunction of the density matrix
with the highest macroscopic eigenvalue \cite{Leggett}.
The local momentum of the condensate depends also on the position and
the time through the relation
\begin{eqnarray}
\vc{k_s}(\vc{r},t)=\nabla_\vc{r} \theta (\vc{r},t)
\end{eqnarray}
Up to a constant phase, the amplitude and the local momentum
of the condensate characterize fully $\Psi(\vc{r},t)$.
In this volume, we define also the local particle number distribution 
or Wigner function 
$n'_{\vc{k}}(\vc{r},t)$ 
(for clarity, explicit dependence of time has been added). 
Also the particles feel an external local potential $V_{eff}(\vc{r})$.
Up to the first order in the interaction potential, the local 
energy density is divided into the kinetic part, an external 
potential part and 
the Hartree-Fock part \cite{Zaremba,BZS}: 
\begin{eqnarray}
{\cal E}(\vc{r},t)=
\frac{|\nabla_{\vc{r}}\Psi(\vc{r},t)|^2}{2m}
+ V_{eff}(\vc{r})|\Psi(\vc{r},t)|^2
\nonumber \\
+\sum_\vc{k} 
\left(\frac{\vc{k}^2}{2m} +V_{eff}(\vc{r})\right)
n'_{\vc{k}}(\vc{r},t)
+\frac{2\pi a}{mV}\big[|\Psi(\vc{r},t)|^4 +
\nonumber \\
4 |\Psi(\vc{r},t)|^2 \sum_{\vc{k}} n'_{\vc{k}}(\vc{r},t)
+2 (\sum_{\vc{k}} n'_{\vc{k}}(\vc{r},t))^2 \big]
\end{eqnarray}
This energy density varies slowly in space. The gradients 
produced by these variations govern the dynamic of the 
gas particle between each small volume $\Omega$. 
The local energy per excited particle with a momentum $\vc{k}$ 
is given by the derivative of the local energy density 
with respect to the particle number \cite{Zaremba}:
\begin{eqnarray}
\epsilon_{\vc{k}}(\vc{r},t)=
\frac{d {\cal E}}{d n'_{\vc{k}}(\vc{r},t)}
=
\frac{\vc{k}^2}{2m}+
{V}_e(\vc{r},t)
\end{eqnarray}
where we define the effective potentials 
felt by the condensate and the particle 
\begin{eqnarray}
{V}_e(\vc{r},t)=
V_{ext}(\vc{r})+
\frac{8\pi a}{mV}
\left[ |\Psi(\vc{r},t)|^2 +
 \sum_{\vc{k}} n'_{\vc{k}}(\vc{r},t) \right]
\end{eqnarray}
Between each infinitesimal volume, particles 
are transferred due to the gradients of the 
particle density and of the potential energy.
As a consequence, the kinetic equations must 
be modified in order to take into account 
locally the modifications inside $\Omega$.
If we consider that the excited particle moves 
classically, then the dynamic of transfer is given 
by the Liouville operator acting on the distribution 
function:
\begin{eqnarray}
\{\epsilon_{\vc{k}}(\vc{r},t),n'_{\vc{k}}(\vc{r},t)\}
\nonumber \\
=\nabla_{\vc{k}}\epsilon_{\vc{k}}(\vc{r},t).
\nabla_{\vc{k}}n'_{\vc{k}}(\vc{r},t)
-\nabla_{\vc{r}}\epsilon_{\vc{k}}(\vc{r},t).
\nabla_{\vc{k}}n'_{\vc{k}}(\vc{r},t)
\end{eqnarray} 
The condensate particles however move locally 
according to the quantum wave function, which 
obeys to the generalized Gross-Pitaevskii 
equation  
with the 
effective potential \cite{Zaremba}: 
\begin{eqnarray}
{V}_c(\vc{r},t)=V_{ext}(\vc{r})+
\frac{4\pi a}{mV}
\left[|\Psi(\vc{r},t)|^2 +
2 \sum_{\vc{k}} n'_{\vc{k}}(\vc{r},t) \right]
\end{eqnarray}

Furthermore, we assume that locally in the
small volume $\Omega$:
1) the system is sufficiently homogeneous so that the expressions 
(\ref{K6}) and (\ref{K7}) remain valid and the condensate is 
not affected by collision; 2) the condensate is
locally macroscopically populated. 
Combining these results together with the collision 
terms, we can write the kinetic equations for 
weakly inhomogeneous and stable Bose gas in the region of condensation. 
We find two coupled set of equations, one 
for the condensate, the other for the normal fluid: 
\begin{widetext}
\begin{eqnarray}\label{inhomo}
i\frac{\partial}{\partial t} \Psi(\vc{r},t)
&=&\left[-\frac{\nabla_\vc{r}^2}{2m}
+{V}_c(\vc{r},t)
\right]\Psi(\vc{r},t)
\\ \label{inhomo2}
\frac{\partial}{\partial t} n'_{\vc{k}}(\vc{r},t)
&=&
\left[-\frac{\vc{k}}{m}.\nabla_\vc{r}
+\nabla_\vc{r}V_{e}(\vc{r},t)
.\nabla_\vc{k}\right]n'_{\vc{k}}(\vc{r},t)
+{\cal C}_{\vc{k}}[n_{\vc{k'}}(\vc{r},t);\vc{k_s}(\vc{r},t)]
\end{eqnarray}
\end{widetext}
With the exception of the collision term, 
the Eq.(\ref{inhomo}) and (\ref{inhomo2}) are identical 
to the kinetic equations formulated by 
Zaremba et al. \cite{Zaremba}. 
The inhomogeneous kinetic equations satisfy  
locally the conservation laws in these regions of highly 
populated condensate. 
We can derive indeed a 
conservation 
equation for the local particle number 
$N(\vc{r},t)=\sum_\vc{k}n_{\vc{k}}(\vc{r},t)$, the 
local momentum $P(\vc{r},t)=\sum_\vc{k}\vc{k} n_{\vc{k}}(\vc{r},t)$
and the total local energy ${\cal E}(\vc{r},t)$ \cite{Zaremba}. It is also 
easy to verify that the production of the local entropy 
$S(\vc{r},t)$ is always positive \cite{Balescu}. 

The local production of entropy stops when we reach 
the local equilibrium:
\begin{eqnarray}\label{neqin}
n'^{eq}_{\vc{k}}(\vc{r},t) =
\frac{1}{\exp{[\beta(\vc{r},t)(\epsilon_\vc{k}-\vc{k}.\vc{v_n}(\vc{r},t)
-\mu(\vc{r},t))]}-1}
\end{eqnarray}
where $\beta(\vc{r},t)$, $\vc{v_n}(\vc{r},t)$ and 
$\mu(\vc{r},t)$ are now local functions of the position and the time.

A gap shows up between the energies of the condensed and non condensed 
particles since 
${V}_c(\vc{r},t) < {V}_e(\vc{r},t)$. This is not a problem 
as long as in the region of the gap the transfer of particle is 
forbidden. The transfer of particle is only possible in the region 
where there is no gap i.e. when the condensate population is not 
macroscopic. 
Therefore, on the 
basis of these non-equilibrium considerations, the Hartree-Fock model 
showing up this forbidden gap does not  
suffer from any inconsistency related to the conservation of energy. 

\subsection{The superfluid universe at finite temperature}

Using the expression (\ref{neqin}) as the solution of 
the kinetic equation, we can find a set of 
equations describing the non dissipative motion of 
the condensed gas at finite temperature. If we assume 
that $\vc{v_n}(\vc{r},t)=0$ then the substitution of 
(\ref{neqin}) in 
Eq.(\ref{inhomo2}) imposes 
that $n'^{eq}_{\vc{k}}(\vc{r},t)$ must be stationary in time, 
that $\beta(\vc{r},t)=\beta$ is a constant and 
that $\mu(\vc{r},t)=-V_e(\vc{r},t)+\mu_e$, where $\mu_e$ is 
a chemical potential independent of the position controlling the number 
of excited particles.
As a consequence, after carrying out 
the integration over the momentum, the local number 
of excited particle is given by the closed equation:
\begin{eqnarray}\label{super1}
N_e^{eq}(\vc{r})=
\sum_{\vc{k}} n'^{eq}_{\vc{k}}(\vc{r}) 
\nonumber \\
=V
\left(\frac{m}{2\pi \beta}\right)^{3/2}
\!\!\!\!
g_{3/2}\left(e^{\beta[
\mu_e -
V_{ext}(\vc{r})-
\frac{8\pi a}{mV}
(|\Psi(\vc{r},t)|^2 +
N_e^{eq}(\vc{r}) )]} \right)
\nonumber \\
\end{eqnarray}
where $g_{l}(x)=\sum_{j=1}^\infty
j^{-l}x^j$. Since the system is stationary, 
the macroscopic condensate wave function has the form:
\begin{eqnarray}
\Psi(\vc{r},t)=
e^{-i\mu_c t}\Psi_c(\vc{r})
\end{eqnarray}
where $\mu_c$ is the chemical potential of 
controlling the population of the condensate particle.  
The substitution of this form into Eq.(\ref{inhomo})
produces:
\begin{eqnarray}\label{super2}
\left[-\frac{\nabla_\vc{r}^2}{2m}
+
V_{ext}(\vc{r})+
\frac{4\pi a}{mV}
(|\Psi_c(\vc{r})|^2 +
2N_e^{eq}(\vc{r}) )
\right]\Psi_c(\vc{r})
\nonumber \\
=
\mu_c \Psi_c(\vc{r})
\end{eqnarray}  
The coupled set of equations 
(\ref{super1}) and (\ref{super2}) describe 
locally all the non dissipative or superfluid 
phenomena. These are valid provided the gas is stable. 
As an example, they describe  
the superfluid moving with a different velocity 
than the normal fluid. In that case when 
$V_{ext}(\vc{r})=0$, the 
solution is  
a plane wave function:
\begin{eqnarray}
\Psi_c(\vc{r})=e^{i\vc{k_s}.\vc{r}}
n_{\vc{k_s}}^{1/2}
\end{eqnarray} 
and $N_e^{eq}(\vc{r})$ is a constant. 
More generally, any non dissipative 
complex structure 
at finite temperature, like vortices, 
should be a solution of (\ref{super1}) and (\ref{super2}).

In contrast to previous 
studies, non-dissipative phenomena are not a 
requirement or assumption of a model but rather a 
prediction of the GRPA kinetic theory. 
Another difference is that the chemical potential 
for both fluids $\mu_c$ and $\mu_e$ must not 
be necessarily identical, contrary to 
the equilibrium statistical mechanics which 
imposes equality \cite{Huang}. In our example, 
$\mu_e=(8\pi a N)/(mV) \not= 
\mu_c=\vc{k_s}^2/(2m) + 4\pi a (2 N_e+n_{\vc{k_s}})/(mV)$ 
at the difference of \cite{Zaremba} where 
$\mu_e=\mu_c$ for $\vc{k_s}=0$.

\section{Analogy with plasmon theory}

The derivation leading to the QKE for a Bose gas 
is similar to 
the one leading to the QKE equation 
for plasma physics \cite{QBalescu, Ichimaru,WP} except 
that, in these equations, the collective excitations 
are derived in the SRPA. In plasma physics, the 
Coulomb interactions 
potential for the collision process is screened 
beyond the Debye wavelength. The dynamic dielectric 
function  displays this screening and removes the 
singularity in the Coulomb potential in the  
long wavelength limit. 
Static screening is well known since the Debye theory. 
However, for dynamic screening, Wyld and Pines have 
proposed an interpretation in terms of the plasmon 
theory. This theory is an alternative version of the kinetic theory 
of a plasma which emphasizes the role played by the collective 
modes. According to their work, the effective potential 
interaction 
is the result of a plasmon mediating the interaction with 
the quasi particles. In other words, during a collision
process, a quasi-particle emits an intermediate plasmon 
on the energy shell with the momentum transfer $\vc{q}$. 
Later on, this plasmon is 
eventually absorded by another quasiparticle which 
acquires a new momentum and a new energy. 
The energy spectrum for this plasmon corresponds  
precisely to the 
frequency spectrum of the 
collective excitations in a plasma.  

In the case of a Bose condensed gas, 
the collective excitations spectrum 
corresponds
to the Bogoliubov frequency spectrum for 
low temperature. This property suggests 
the interpretation that these phonon-like 
excitations play also the role of 
mediators during collisions between 
quasi-particles. Thus, the Bose condensation 
has the effect to transform the scattering modes, 
used for collision involving condensed particle, 
into a collective 
mode used for mediating the interaction between 
non condensed particles.

The plasmon theory has been derived using 
the theory of quantum 
electrodynamics. In principle, 
for a Bose gas, a similar 
approach must be carried out taking into account 
that, at a more fundamental level, 
the interaction potential originates 
from processes involving the absorption 
and the emission of photons.
In this paper, we shall not rederive 
a similar theory but rather recover 
heuristically 
the analogy 
with the plasmon theory.  

Following this approach, this intermediate
excitation 
behaves like a particle characterized by its 
own distribution function $f_{\vc{q}}$ and that 
we shall call by analogy "condenson".
The energy spectrum of the condenson is given by 
the zeroes of the dynamic dielectric function i.e. 
by the solution of $\Delta(\vc{q},\omega)=0$. 
We consider the simple 
case of an homogenous condensate at rest $\vc{k_s}=0$ 
and weakly depleted.
In that case,  
the solution is the complex number 
$\omega=\omega_{\vc{q}}-i\gamma_{\vc{q}}$, 
where the real part represents the energy
spectrum $\omega_{\vc{q}}=
\epsilon_{\vc{q}}^B$
and the imaginary part represents the decay 
rate of the condenson given by  Eq.(\ref{imo}).
Then, looking at Eq.(7) and Eq.(8) in \cite{WP}, we can 
write two coupled equations, one for the dynamic 
evolution 
of the quasi-particle and the other 
for the evolution of the condenson. 
They describe the time rate change of these quantities 
due to the emission and absorption of a condenson by a 
quasi-particle:
\begin{widetext}
\begin{eqnarray}\label{WP1}
\frac{\partial}{\partial t} n'_{\vc{k}}
&=&\frac{8\pi a}{mV}\!\! \sum_{\vc{q}}
\frac{8\pi a n_\vc{0} \vc{q}^2}{m^2 V\omega_\vc{q}}
\big[
\pi \delta(\omega_{\vc{q}}+
\epsilon_{\vc{k}}-\epsilon_{\vc{k}+\vc{q}})
\left(
(n'_{\vc{k}}+1)n'_{\vc{k}+\vc{q}}
(f_{\vc{q}}+1)
-n'_{\vc{k}}(n'_{\vc{k}+\vc{q}}+1)f_{\vc{q}}
\right)
\nonumber \\
&& +\pi\delta(\omega_{\vc{q}}-
\epsilon_{\vc{k}}+\epsilon_{\vc{k}-\vc{q}})
\left(
(n'_{\vc{k}}+1)n'_{\vc{k}-\vc{q}}f_{\vc{q}}
-n'_{\vc{k}}(n'_{\vc{k}-\vc{q}}+1)(f_{\vc{q}}+1)
\right)
\big]
\\ \label{WP2}
\frac{\partial}{\partial t}f_{\vc{q}}&=&
-2\gamma_{\vc{q}}f_{\vc{q}}
+\frac{8\pi a }{mV}
\frac{8\pi a n_\vc{0} \vc{q}^2}{m^2V\omega_\vc{q}}
\sum_{\vc{k}}\pi\delta(\omega_{\vc{q}}-
\epsilon_{\vc{k}}+\epsilon_{\vc{k}-\vc{q}})
n'_{\vc{k}}(n'_{\vc{k}-\vc{q}}+1)
\end{eqnarray}
\end{widetext}
Similarly to the reasoning of Wyld and Pines,
the kinetic equation (\ref{K6}) can be rederived
from Eq.(\ref{WP1}) and Eq.(\ref{WP2}) by eliminating 
adiabatically $f_\vc{q}$. We assume that 
the time
derivative in Eq.(\ref{WP2}) is zero
and the elimination allows to obtain an equation 
for $n'_{\vc{k}}$ only. Eq.(\ref{K6}) is recovered, 
provided  that
the condensate is weakly depleted. Under these circonstances, 
the square of the effective potential has a narrow Lorentzian  
shape whose two peaks and widths correspond to the condenson 
energy and decay rate respectively (see \cite{WP}). It 
plays the role of a propagator for the mediated interaction.
This supposes that the dielectric function can be 
approximated around the resonant frequencies as 
(see Eq.(25) in \cite{WP} for comparison):
\begin{eqnarray}\label{ka}
{\cal K}(\vc{q},\omega)=\left\{
\begin{array}{ccc}
\frac{m^2 V\omega_{\vc{q}}}{4\pi a n_\vc{0} \vc{q}^2}
(\omega -\omega_{\vc{q}}+i\gamma_{\vc{q}}),& 
\omega \sim \omega_{\vc{q}}\\
 -\frac{m^2 V\omega_{-\vc{q}}}{4\pi a n_\vc{0} \vc{q}^2}
(\omega + \omega_{-\vc{q}}-i\gamma_{-\vc{q}}),&
\omega \sim -\omega_{-\vc{q}}
\end{array}
\right.
\end{eqnarray}
leading to the approximation:
\begin{eqnarray}
\frac{1}{|{\cal K}(\vc{q},\omega)|^2}
=
\frac{\pi}{\gamma_{\vc{q}}}
\left(\frac{4\pi a n_\vc{0} \vc{q}^2}{m^2 V\omega_\vc{q}}\right)^2
\!\!\!\!\left(
\delta(\omega-\omega_{\vc{q}})+
\delta(\omega+\omega_{\vc{q}})
\right)
\end{eqnarray}
These coupled equations suggest that a weakly interacting  
Bose 
condensed gas is in reality composed of two kind of 
bosonic particles:
the quasi-particles that become the real particles in 
the limit of an ideal Bose gas and the condensons that 
result from a mediation (see Fig.1).
At thermal equilibrium, the various particles 
obey to the Bose-Einstein statistics: 
\begin{eqnarray}\label{neq2}
n'_{\vc{k}}=n^{eq}_{\vc{k}}=
\frac{1}{\exp{[\beta(\epsilon_\vc{k}-\mu)]}-1}
\end{eqnarray}
and 
\begin{eqnarray}\label{feq}
f_{\vc{q}}=
\frac{1}{\exp{(\beta \omega_{\vc{q}})}-1}
\end{eqnarray}
In this way, the condenson distribution function 
has the same form as the  
excitations distribution function
predicted by Bogoliubov \cite{Pines}. 
But there is an important difference however:

{\it 
In the Bogoliubov theory, the Bogoliubov 
excitations 
correspond to the quasi-particle. On the contrary, in the GRPA 
model, the Bogoliubov excitations correspond to the 
condensons and, in this respect, are not the 
quasi-particles.}

The QKEs proposed by \cite{Imamovic,Walser,Kirkpatrick,Gardiner}
suggest instead that 
the Bogoliubov excitations correspond to the quasi-particle.  
We must be cautious with this interpretation since 
the GRPA model is not an exact one. It might be that, 
to a next order expansion in the interaction, 
the quasi-particles have a different energy spectrum that is linear 
in the momentum. Without any further investigation, 
we cannot claim that the Goldstone boson, that would be 
normally predicted from the $U(1)$ theory, 
is the condenson or the quasi-particle. 

\begin{figure}
\scalebox{1}{
\includegraphics{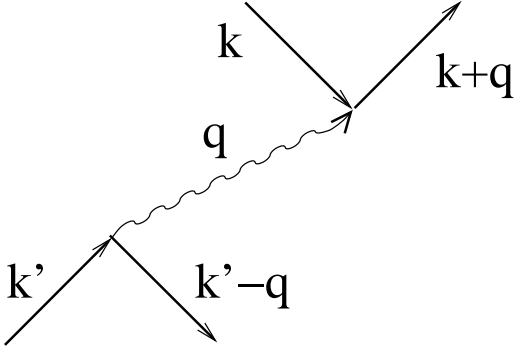}} 
\caption{Feymann diagram of the 
interaction of two quasi-particle (full line)
mediated by the condenson (wavy line)}  
\end{figure}

The condenson has a decay rate $2\gamma_{\vc{q}}$. 
Thus, a duration of the collision is limited 
by its lifetime and can be considered as instantaneous, 
if much lower than the inverse relaxation rate. 
Assuming that the particle has a mean velocity 
$1/(\beta m)^{1/2}$, this requirement imposes for the 
typical values:
$\gamma_{\vc{q}} \gg \rho_e \sigma /(\beta m)^{1/2}$.
If it is not fulfilled, then the long lifetime 
of the condenson suggests that its dynamic evolution 
is not instantaneous and that the full set of equations 
(\ref{WP1}) and (\ref{WP2}) must be rather used.

\section{Equivalence with the Bogoliubov theory}

The RPA model presents important  differences compared to
any previous model for finite temperature.
However for zero temperature, it
allows to recover the next order correction for the ground state
energy and the static structure factor 
usually obtained from the Bogoliubov theory.
Such a derivation has already been made by
Nozi\`eres and Pines for an electron gas
using the density response formalism.
They were also able to apply
this formalism to the condensate \cite{Pines}.
The nice result of this section is that,
precisely, the formalism developed in this
paper allows to recover exactly the same results for the 
ground state energy and the structure factor.

Indeed, the correlation function defined in (\ref{g}) allows
to calculate the correction to the energy but also to 
the total particle momentum
distribution $N_{\vc{k}}$. This new distribution differs 
from the quasi-particles population 
distribution $n_{\vc{k}}$. We start
initially from the distribution without interaction
$N_{\vc{k}}=\delta_{\vc{k},\vc{0}} n_{\vc{0}}$ where
only the condensed mode $\vc{k_s}=0$ is populated. Then,
we switch on adiabatically the interaction and the
correlations appear in a form given by (\ref{K3}).
As a consequence, the condensate becomes weakly depleted
and the total kinetic and potential energy gets modified.

These corrections can be
calculated directly the static structure factor
which, using (\ref{g}) and for $\vc{q} \not= 0$, 
is given by:
\begin{eqnarray}\label{SF}
N S(\vc{q})=\langle \rho^\dagger_{\vc{q}}\rho_{\vc{q}}
\rangle=
\sum_{\vc{k}}(n_{\vc{k}}+1)n_{\vc{k}+\vc{q}}
+
\sum_{\vc{k},\vc{k'}}
g_\vc{q}(\vc{k},\vc{k'}) 
\end{eqnarray}
where $N=\sum_{\vc{k}}N_{\vc{k}}=\sum_{\vc{k}} n_{\vc{k}}$. 
An expression for this integral can be found
in the appendix C. 
In contrast, Nozi\`eres and Pines express the 
structure factor in terms of the susceptibility function 
(\ref{susc}) of the ground state 
as \cite{NP}:
\begin{eqnarray}\label{SFNP}
NS(\vc{q})=\langle \rho^\dagger_{\vc{q}}\rho_{\vc{q}}
\rangle =-
\int_0^\infty
\frac{d\omega}{\pi}
{\rm Im} \chi (\vc{q},\omega)
\end{eqnarray}
At zero temperature, the limit
$n'_{\vc{k}} \rightarrow 0$ must be taken carefully.
It implies that the Landau damping
$\gamma_{\vc{q}} \rightarrow 0$ but this parameter is
different from $0_+$. The limit $0_+ \rightarrow 0$
must be carried out before the limit
$\gamma_{\vc{q}} \rightarrow 0$ and this procedure
can be done by adjusting adequately the infinitesimal value of $0_+$
and $n'_{\vc{k}}$. In particular, for a normal fluid 
in equilibrium,  this amounts to prescribing to 
take the limit $0_+$ to zero and then after taking  the 
limit $\beta \rightarrow \infty$ in Eq.(\ref{neq}).
The physical reason for such
a procedure is that the presence of infinitesimal fraction
of excited particles will force the system to create
the correct equilibrium correlation. 
In these conditions and
using Eq.(\ref{Knapp}) and 
Eq.(\ref{imo}), we can re express for $\vc{k_s}=0$:
\begin{eqnarray}\label{disprel1}
\Delta(\vc{q},\omega) \rightarrow
(\omega+i\gamma_{\vc{q}})^2-
{\epsilon^B_{\vc{q}}}^2
\end{eqnarray}
Also, we note the relation: 
\begin{eqnarray}
{\rm Im}A'(\vc{q},\omega)=
-\frac{mV}{8\pi a}
\frac{1}{\exp(\beta \omega)-1}
{\rm Im}{\cal K}_n (\vc{q},\omega)
\end{eqnarray}
Inserting these results 
and (\ref{imo}) into
(\ref{intg}), setting to zero the real part
of ${\cal K}(\vc{q},\omega)$ and $A'(\vc{q},\omega)$, and 
setting $A_0 (\vc{q},\omega)=N/[\omega+i0^+ +\vc{q}^2/(2m)]$ 
then Eq.(\ref{SF}) becomes:
\begin{eqnarray}
S(\vc{q})
&=&1 - \int_{-\infty}^\infty
\frac{d\omega}{\pi}
\frac{\omega^2-\left(\frac{\vc{q}^2}{2m}\right)^2 +
\frac{4\pi a N}{mV}\frac{\vc{q}^2}{m}}
{|(\omega+i\gamma_{\vc{q}})^2-
{\epsilon^B_{\vc{q}}}^2|^2}
\nonumber \\ 
\nonumber && \times
\left(\omega-\frac{\vc{q}^2}{2m}
\frac{e^{\beta \omega}+1}{e^{\beta \omega}-1}\right)  
{\rm Im}{\cal K}_n (\vc{q},\omega)\\
\label{T0'}
&\stackrel{\begin{array}{c} \gamma_{\vc{q}} \rightarrow 0
\\ \beta \rightarrow \infty \end{array}}{=}&
\frac{\vc{q}^2}{2m\epsilon^B_{\vc{q}}}
\end{eqnarray}
In order to get
(\ref{T0'}),
the integration has been carried out transforming 
the Lorentzian factors into delta functions and using 
(\ref{imo}). This result is identical to that
obtained from the Bogoliubov theory.
This result can also been obtained substituting (\ref{susc}) 
into (\ref{SFNP}) and carry out
the integration over $\omega$.
Instead of evaluating the total energy and the 
correction to the particle number distribution directly,
some useful tricks   
allow to 
avoid some complex calculation \cite{NP}. The total ground state 
energy is  
a functional of the interaction $a$ and the dispersion relation 
$\epsilon_{\vc{k}}$ (in what follows, we consider only 
the first term in (\ref{U0})). It can be expressed in terms of 
the matrix element
$E(a,\epsilon_{\vc{k}})=
\langle \psi| H |\psi \rangle$
where $|\psi \rangle$ is 
the ground state depending also on $a$ and $\epsilon_{\vc{k}}$. 
Using the property that $|\psi \rangle$ is a normalized 
eigenfunction,
the first derivative with respect to these parameters 
produces:
\begin{eqnarray}\label{da}
a \frac{\partial E(a,\epsilon_{\vc{k}})}
{\partial a}&=& E_{int}(a,\epsilon_{\vc{k}})
\\ \label{de}
\frac{\partial E(a,\epsilon_{\vc{k}})}
{\partial \epsilon_{\vc{k}}}&=& N_{\vc{k}}
\end{eqnarray}
The total interaction energy can be expressed in terms 
of the structure factor:
\begin{eqnarray}\label{T01}
E_{int}(a,\epsilon_{\vc{k}})=
\frac{2\pi a}{{mV}}\left[ N(N-1)+N\sum_{\vc{q}\not=0}
(S(\vc{q})-1)\right]
\end{eqnarray}
At zero temperature and without interaction, 
all the particles are in the ground state and thus 
$E(0,\epsilon_{\vc{k}})=N\epsilon_{\vc{0}}$.
With this initial condition,  
an integration over $a$ allows to calculate the total energy:
\begin{eqnarray}\label{T03}
E(a,\epsilon_{\vc{k}})=N\epsilon_{\vc{0}}+ 
\int_0^a \frac{da'}{a'} E_{int}(a',\epsilon_{\vc{k}})
\nonumber \\ +
(\frac{2\pi a N}{V})^2\sum_{\vc{q}\not=0}\frac{2m}{\vc{q}^2}
\end{eqnarray}
where a supplementary term coming from (\ref{U0}) 
has been added to 
remove ultra-violet divergencies. The structure factor 
can be recalculated for any arbitrary dispersion relation 
$\epsilon_{\vc{k}}$. Plugging this new expression into 
(\ref{T01}) and carrying out the integral over $a$, we obtain:
\begin{eqnarray}\label{T04}
E(a,\epsilon_{\vc{k}})=
N\epsilon_{\vc{0}}+
\frac{2\pi a N^2}{mV}+
\frac{1}{2}
\sum_{\vc{q}\not=0} \bigg[
(\frac{2\pi a N}{V})^2\frac{m}{\vc{q}^2}+
\nonumber \\
\sqrt{\Delta\epsilon_{\vc{q}}^2+\frac{8\pi a N}{mV}
\Delta\epsilon_{\vc{q}}}-\Delta\epsilon_{\vc{q}}
-\frac{4\pi a N}{mV} \bigg]
\end{eqnarray}
where $\Delta\epsilon_{\vc{q}}=
(\epsilon_{\vc{q}}+\epsilon_{-\vc{q}}-
2\epsilon_{\vc{0}})/2$. In the usual case, $\Delta\epsilon_{\vc{q}}=
\vc{q}^2/2m$ and an integration over the momentum $\vc{q}$ allows 
to find the ground state energy \cite{Huang}:
\begin{eqnarray}\label{Ecor}
E(a)=
\frac{2\pi a N^2}{mV}
[1+ \frac{128}{15}\sqrt{\frac{a^3 N}{\pi V}}]
\end{eqnarray}
Finally, the derivative with respect 
to $\epsilon_{\vc{k}}$ gives the first 
correction to the momentum particle distribution. We get
for $\vc{k}\not= 0$:
\begin{eqnarray}\label{ncor}
N_{\vc{k}}=\left. \frac{\partial E(a,\epsilon_{\vc{k}})}
{\partial \epsilon_{\vc{k}}}\right|_{
\epsilon_{\vc{k}}=\frac{\vc{k}^2}{2m}} 
=\frac{1}{2}\left(\frac{
\frac{\vc{k}^2}{2m}+\frac{4\pi a N}{mV}}{
\epsilon^B_{\vc{k}}}-1\right)
\end{eqnarray}
and for $\vc{k}=0$:
\begin{eqnarray}
N_{\vc{0}}=N- \sum_{\vc{k}\not=0}N_{\vc{k}}
=N[1-\frac{32}{3}\sqrt{\frac{a^3 N}{\pi V}}]
\end{eqnarray}
From our number conserving approach, we recover the well-known
results usually obtained from the Bogoliubov approach at 
zero temperature, 
i.e. the first order energy correction in (\ref{Ecor}) 
and the first order 
correction to the number of excited particles (\ref{ncor}) showing 
a depletion of the condensate particle. 
In principle, the present method can be extended to 
finite temperature, since the relations (\ref{da}) and 
(\ref{de}) are valid also for any excited state labeled 
by the occupation number $n_{\vc{k}}$. The only problem 
remains the explicit calculation of 
$\langle \rho^\dagger_{\vc{q}}\rho_{\vc{q}}
\rangle$.


\section{Conclusions and Perspectives}

We attempted to reexamine the kinetic theory of 
the weakly interacting and stable Bose condensed gas with the 
objective to explain superfluidity for any temperature 
and, ultimately, to 
be able to distinguish between its dissipative and 
superfluid behaviors. A different 
QKE, taking into account some higher order terms 
in the interaction parameter, has been derived from the 
microscopic theory 
and has the merit to predict a collision blockade 
between condensed and non condensed particle. 
{\it More precisely, the condensate remains invisible to any
thermal quasi-particle due to an induced force, which
acts as a `` coarse-graining '', removing any
local force necessary for binary scattering.}
In this way a metastable state can be built 
locally with a non-zero relative velocity. 
For the homogeneous gas, a microscopic derivation 
has been carried out using the same expansion procedure as that  
for deriving the QKE for a quantum plasma. The only 
difference is that our approach takes into account the 
exchange term which is of the same order of magnitude 
than the direct term.
For the weakly inhomogeneous 
gas, the derivation needs some supplementary 
justifications from first principles, in particular, for the way 
to obtain the generalized Gross-Pitaevskii 
equation from a number conserving approach, but also for the 
limit of use of the collision terms (\ref{K6}) and (\ref{K7}).

The generalized RPA is the basic approximation 
which is at the origin of the prediction of the collision 
blockade phenomenon. 
It is not only a method to achieve 
results but has also the deep meaning that average 
contributions that are not oscillating in phase can 
be neglected for diluted gas. The RPA does not serve 
only to predict the QKE; it allows to build a genuine 
alternative theory that 
can be compared with all previous  approaches. The  
main advantages of this theory are:
1) it is
number conserving; 2) the statistical
equilibrium is not imposed as a postulate but is deduced 
from the QKE  
and allows superfluidity; 3) it is valid for any 
range of temperature below and above the transition point.

The predictions calculated 
from this theory, namely the collective excitations, 
could be compared with results 
predicted 
mainly by the Popov theory 
or by the Hartree-Fock-Bogoliubov theory in the Popov 
approximation \cite{Popov,Griffin}. 
The RPA theory 
allows also to determine the correlation function 
$g_\vc{q}(\vc{k},\vc{k'})$ with which one can calculate 
the next order correction. In particular, we recover the 
Bogoliubov results for the next correction to the ground
state energy and for the particle momentum distribution. 
In principle, these corrections  
can also be determined at finite temperature 
and compared with other previous works. Another interesting 
quantity is the static structure factor $S(\vc{q})$ 
which can be evaluated directly from the correlation function. 

The RPA theory for dilute Bose gas has some resemblance 
with the plasma kinetic theory. Both 
predict that scattering proceeds by means of an 
intermediate excitation.
For a plasma,  this excitation is a plasmon responsible 
for the collective plasma oscillation. For a Bose gas, 
it is a phonon-like excitation with a Bogoliubov energy 
spectrum at zero temperature and that, by analogy, we can 
call ``condenson''.

Let us also mention the limitation of the present theory.
Firstly, this theory preserves the existence of a gap 
in apparent contradiction with the Hugenholtz-Pines theorem \cite{Pines}.
We consider only the Hartree-Fock approximation for the 
potential energy of
the quasi-particle, excluding in this way higher order 
correlated term.
This contradiction might be solved by claiming 
that the poles of the one particle Green function have a
richer structure; namely one gapless pole which represents 
the Bogoliubov or condenson excitation spectrum and another pole 
which may have a gap and which represents the quasi-particle 
spectrum. In this way, we recover the compatibility with the 
HP theorem.
Secondly, in principle, the calculation must start with the 
real potential and not the effective contact 
potential (\ref{U0}). A resummation of all ladder diagrams 
should allow to reexpress the effective interaction in terms 
of the $T$ matrix for the binary collision. Then the limit 
of low energy allows to deduce the expression (\ref{U0}) \cite{Stoof}.
If rather the $T$ many body matrix is considered, then the 
effective interaction could depend also non linearly on 
the quasi-particle number. 
Thirdly, the assumption that the gas is weakly homogeneous might 
not be true, due to a sharp trap potential realized in real 
experiment. The theory must be improved by decomposing the 
one particle Wigner function in terms of its eigenfunctions  
and by deriving an equation of motion for each of them \cite{Leggett}.
  
Finally, all the predictions, that the RPA theory might suggest, 
must be ultimately confronted to experiments in order to be validated.
In particular, the analysis of the various modes for 
the collective excitations and its damping must be recovered 
\cite{JZ}. 

\appendix

\section{Solution of the equation of motion for the excitations}

The system (\ref{col1}) (\ref{col2}) and (\ref{col3}) can be solved
exactly. We isolate the dielectric propagator on the left 
hand side of each equation. After summing over the superfluid and 
normal modes separately and using (\ref{kn}), 
we obtain after rearranging:
\begin{widetext}
\begin{eqnarray}\label{col12}
[(\omega+i0_+ -
\frac{\vc{k_s}.\vc{q}}{m})^2-{\epsilon^B_{\vc{q}}}^2]
\left({\cal U}_\vc{q}(\vc{k_s},\vc{k_1},\omega) +
{\cal U}_\vc{q}(\vc{k_s}-\vc{q} ,\vc{k_1},\omega)
\right)-\frac{8\pi a n_{\vc{k_s}}}{mV}
\frac{\vc{q}^2}{m}\sum'_{\vc{k'}}
\tilde{{\cal U}}_\vc{q}(\vc{k'},\vc{k_1},\omega)
=
\nonumber \\
i
[\omega+i0_+ -
\frac{\vc{k_s}.\vc{q}}{m}+\frac{\vc{q}^2}{2m}]
\delta_{\vc{k_s},\vc{k_1}}
+i
[\omega+i0_+ -
\frac{\vc{k_s}.\vc{q}}{m}-\frac{\vc{q}^2}{2m}]
\delta_{\vc{k_s}-\vc{q},\vc{k_1}}
\end{eqnarray}
\begin{eqnarray}\label{col32}
\left({\cal K}_n(\vc{q},\omega)-1
\right)
\left({\cal U}_\vc{q}(\vc{k_s},\vc{k_1},\omega) +
{\cal U}_\vc{q}(\vc{k_s}-\vc{q} ,\vc{k_1},\omega)
\right)
+
{\cal K}_n(\vc{q},\omega)
\sum'_{\vc{k'}}
\tilde{{\cal U}}_\vc{q}(\vc{k'},\vc{k_1},\omega)
=
i\frac{1-\Pi_{\vc{k_1}}
}{\omega+i0_+ -
\frac{\vc{k_1}.\vc{q}}{m}-\frac{\vc{q}^2}{2m}}
\end{eqnarray}
where we define the function 
$\Pi_{\vc{k}}=\delta_{\vc{k_s},\vc{k}}+
\delta_{\vc{k_s}-\vc{q},\vc{k}}$. Using the definition 
(\ref{disprel}),
the solution of this coupled set of equations gives:
\begin{eqnarray}\label{col13}
{\cal U}_\vc{q}(\vc{k_s},\vc{k_1},\omega) +
{\cal U}_\vc{q}(\vc{k_s}-\vc{q} ,\vc{k_1},\omega)
=
i\frac{{\cal K}_n(\vc{q},\omega)\Pi_{\vc{k_1}}
\left[(\omega+i0_+ -
\frac{\vc{k_s}.\vc{q}}{m})^2-(\frac{\vc{q}^2}{2m})^2\right]
+\frac{8\pi a n_{\vc{k_s}}}{mV}\frac{\vc{q}^2}{m}
(1-\Pi_{\vc{k_1}})
}{\Delta(\vc{q},\omega)
[\omega+i0_+ -
\frac{\vc{k_1}.\vc{q}}{m}-\frac{\vc{q}^2}{2m}]}
\end{eqnarray}
and 
\begin{eqnarray}\label{col33}
\sum'_{\vc{k'}}
\tilde{{\cal U}}_\vc{q}(\vc{k'},\vc{k_1},\omega)
=
i\frac{(1-{\cal K}_n(\vc{q},\omega))
\left[(\omega+i0_+ -
\frac{\vc{k_s}.\vc{q}}{m})^2-(\frac{\vc{q}^2}{2m})^2\right]
\Pi_{\vc{k_1}}+
[(\omega+i0_+ -
\frac{\vc{k_s}.\vc{q}}{m})^2-{\epsilon^B_{\vc{q}}}^2]
(1-\Pi_{\vc{k_1}})
}{\Delta(\vc{q},\omega)
[\omega+i0_+ -
\frac{\vc{k_1}.\vc{q}}{m}-\frac{\vc{q}^2}{2m}]}
\end{eqnarray}
Consequently, the definitions (\ref{kalc}) and 
(\ref{kalc2}) allow to write
\begin{eqnarray}\label{col34}
\sum_{\vc{k'}}
{\cal U}_\vc{q}(\vc{k'},\vc{k_1},\omega)
=\frac{i}{\omega+i0_+ -
\frac{\vc{k_1}.\vc{q}}{m}-\frac{\vc{q}^2}{2m}}
\left(\frac{1-\Pi_{\vc{k_1}}}{{\cal K}(\vc{q},\omega)}
+\frac{\Pi_{\vc{k_1}}}{{\tilde {\cal K}}(\vc{q},\omega)}\right)
\end{eqnarray}
Plugging these results into the r.h.s. of
(\ref{col1}) (\ref{col2}) and (\ref{col3}), the integral 
terms do not appear anymore and we get the solution:
\begin{eqnarray}\label{Ur}
{\cal U}_\vc{q}(\vc{k},\vc{k_1},\omega)
=\frac{i}{
\omega+i0_+ -
\frac{\vc{k}.\vc{q}}{m}-\frac{\vc{q}^2}{2m}}
\left[\delta_{\vc{k},\vc{k_1}}+
\frac{
\frac{8\pi a }{mV}
(n_{\vc{k}}-n_{\vc{k+q}}){\cal K}^{-1}_{\vc{k},\vc{k_1}}(\vc{q},\omega)}
{
\omega+i0_+ -
\frac{\vc{k_1}.\vc{q}}{m}-\frac{\vc{q}^2}{2m}}
\right]
\end{eqnarray}
where we define the reverse dielectric function
\begin{eqnarray}\label{reverse}
{\cal K}^{-1}_{\vc{k},\vc{k_1}}(\vc{q},\omega)
=
\frac{(1-\Pi_{\vc{k}})(1-\Pi_{\vc{k_1}})}
{{\cal K}(\vc{q},\omega)}
+
\frac{(1-\Pi_{\vc{k}})\Pi_{\vc{k_1}}+
\Pi_{\vc{k}}(1-\Pi_{\vc{k_1}})+
\Pi_{\vc{k}}\Pi_{\vc{k_1}}\left(1-{\cal K}_n(\vc{q},\omega)/2\right)}
{{\tilde {\cal K}}(\vc{q},\omega)}
\end{eqnarray}
This function can be rewritten in matrix notation, 
where we distinguish in the column and the row 
the channel $\vc{k_s}$ and $\vc{k_s}-\vc{q}$ from the 
other channels:
\begin{eqnarray}\label{reversem}
{\cal K}^{-1}_{\vc{k},\vc{k_1}}(\vc{q},\omega)
\equiv
\left(
\begin{array}{cc}
{\cal K}^{-1}(\vc{q},\omega) & {\tilde {\cal K}}^{-1}(\vc{q},\omega) \\
{\tilde {\cal K}}^{-1}(\vc{q},\omega) & 
(1-\frac{{\cal K}_n(\vc{q},\omega)}{2})
{\tilde {\cal K}}^{-1}(\vc{q},\omega) 
\end{array}
\right)
\end{eqnarray}

\end{widetext}

\section{Calculation of the normal 
dielectric function at equilibrium}

The expression (\ref{kn}) can be put in the form 
of an integral over $t$:
\begin{eqnarray}\label{kn1}
{\cal K}_n(\vc{q},\omega)-1= 
\nonumber \\
i\frac{8\pi a }{mV}\sum_{\vc{k}}
\int_0^\infty \!\!\!dt \,
e^{i(\omega+i0_+ -
\frac{\vc{k}.\vc{q}}{m}-\frac{\vc{q}^2}{2m})t}
(n'^{eq}_{\vc{k}}-n'^{eq}_{\vc{k+q}})
\end{eqnarray}
Using the development:
\begin{eqnarray}\label{kn2}
n'^{eq}_{\vc{k}}=
\sum_{j=1}^\infty
e^{-\beta j(\frac{\vc{k}^2}{2m}-\mu)}
\end{eqnarray}
we can carry out the integration over the 
momentum $\vc{k}$ to get:
\begin{eqnarray}\label{kn3}
{\cal K}_n(\vc{q},\omega)=
1+ \frac{16\pi a }{m}
\int_0^\infty \!\!\!dt\,
e^{i(\omega+i0_+)t}
\sin(\frac{\vc{q}^2}{2m}t)
\nonumber \\
\sum_{j=1}^\infty
\left(\frac{m}{2 \pi \beta j}\right)^{3/2}
e^{\beta j \mu-\frac{\vc{q}^2 t^2}{2\beta j m}}
\end{eqnarray}
As far as the imaginary part is concerned 
the integral over the time can be done 
analytically. The result is:
\begin{eqnarray}\label{kn4}
{\rm Im}{\cal K}_n(\vc{q},\omega)
=
\frac{2a m}{\beta |\vc{q}|}
\ln\left(\frac{1-e^{-\beta[
m\frac{(\omega+\frac{\vc{q}^2}{2m})^2}{2\vc{q}^2}
-\mu]}}
{1-e^{-\beta[
m\frac{(\omega-\frac{\vc{q}^2}{2m})^2}{2\vc{q}^2}
-\mu]}}\right)
\end{eqnarray}
For the real part the expression, (\ref{kn3}) 
needs to be developed in series. For  low temperature, 
${\rm Re}{\cal K}_n(\vc{q},\omega) \simeq 1$.
These results are in agreement with \cite{SK}
but with the difference that we have included 
the exchange term which doubles the interaction 
strength.  
	
\section{Calculation of the collision term}

As said in section 3, the calculation in the SRPA is 
easy since it presents strong resemblance with the plasma gas.  
The extension
to GRPA is more complicated  because now the condensed and 
normal modes cannot be treated on the same foot anymore, 
as long as the interaction energy per particle is not 
the same. Nevertheless, the method developed by 
Ichimaru for a classical plasma can be adapted 
straightforwardly. Here we shall give 
the intermediate steps of the calculations. 

Before doing so, few symmetry properties are interesting:
\begin{eqnarray}
g_\vc{q}(\vc{k},\vc{k'})&=&g^*_\vc{q}(\vc{k-q},\vc{k'+q})
\\
\left({\cal K}^{-1}_{\vc{k},\vc{k_1}}(\vc{q},\omega)\right)^*
&=&{\cal K}^{-1}_{\vc{k},\vc{k_1}}(-\vc{q},-\omega)
\end{eqnarray}
The first  property can be checked from the definition 
(\ref{g}) and the third one is straightforward.
Using the first property and (\ref{n2}),
the collision term defined in Eq.(\ref{K5}) can be expressed in terms 
of the imaginary part of the correlation function:
\begin{eqnarray}\label{B1}
{\cal C}^T_{\vc{k}}[n_{\vc{k'}};\vc{k_s}]=
\sum_{\vc{q},\vc{k'}}\frac{4\pi a}{mV}
{\rm Im}(g_\vc{q}(\vc{k},\vc{k'})
-g_\vc{q}(\vc{k}-\vc{q},\vc{k'}))
\end{eqnarray}
Let us define:
\begin{eqnarray}
A'(\vc{q},\omega)=
\sum_{\vc{k}}(1-\Pi_{\vc{k}})
\frac{(n_{\vc{k}}+1)n_{\vc{k}+\vc{q}}}
{\omega+i0^+ -\frac{\vc{k}.\vc{q}}{m}-\frac{\vc{q}^2}{2m}}
\end{eqnarray}
\begin{eqnarray}
A_0(\vc{q},\omega)
=\sum_{\vc{k}} \Pi_{\vc{k}}
\frac{(n_{\vc{k}}+1)n_{\vc{k}+\vc{q}}}
{\omega+i0^+ -\frac{\vc{k}.\vc{q}}{m}-\frac{\vc{q}^2}{2m}}
\end{eqnarray}
With these definitions together with  
(\ref{col34}), (\ref{Ur}) and (\ref{Q}), carrying out the sum over 
$\vc{k_1}$ and $\vc{k'_1}$ and taking the thermodynamic limit,
after a 
straightforward but lengthly calculation we find the 
following expressions for (\ref{K4}):
\begin{widetext}
\begin{eqnarray}\label{B2}
\sum_{\vc{q},\vc{k'}}(1- \Pi_\vc{k})
g_\vc{q}(\vc{k},\vc{k'})
=
-\sum_{\vc{q}}
\int_{-\infty}^\infty
\frac{d\omega}{2\pi i}
\frac{1}{\omega+i0_+ -
\frac{\vc{k}.\vc{q}}{m}-\frac{\vc{q}^2}{2m}}
\bigg\{
(n_{\vc{k}}+1)n_{\vc{k}+\vc{q}}
\left(\frac{1}{{{\cal K}^*(\vc{q},\omega)}}-1 \right)
\nonumber \\
+
\frac{8\pi a }{mV}(n_{\vc{k}}-n_{\vc{k}+\vc{q}})
\bigg[\frac{(1-{{\cal K}^*(\vc{q},\omega)})A'(\vc{q},\omega)
-A'^*(\vc{q},\omega)}{|{{\cal K}(\vc{q},\omega)}|^2}
+
\frac{A_0(\vc{q},\omega)(1-{\cal K}^*(\vc{q},\omega))-
A^*_0(\vc{q},\omega)}
{|{\tilde{\cal K}}(\vc{q},\omega)|^2}
\bigg]
\bigg\}
\end{eqnarray}

\begin{eqnarray}\label{B3}
\sum_{\vc{q},\vc{k'}}
g_\vc{q}(\vc{k_s},\vc{k'})
=
-\sum_{\vc{q}}
\int_{-\infty}^\infty
\frac{d\omega}{2\pi i}
\frac{1}{\omega+i0_+ -
\frac{\vc{k_s}.\vc{q}}{m}-\frac{\vc{q}^2}{2m}}
\bigg\{
(n_{\vc{k_s}}+1)n_{\vc{k_s}+\vc{q}}
\left(\frac{1}{{\tilde{{\cal K}}^*(\vc{q},\omega)}}-1 \right)
\nonumber \\
+
\frac{8\pi a }{mV}(n_{\vc{k_s}}-n_{\vc{k_s+q}})
\bigg[\frac{(1-{{\cal K}^*(\vc{q},\omega)})A'(\vc{q},\omega)
-A'^*(\vc{q},\omega)}{{\tilde{{\cal K}}(\vc{q},\omega)}
{{\cal K}^*(\vc{q},\omega)}}
+
(1-\frac{{\cal K}_n(\vc{q},\omega)}{2})
\frac{A_0(\vc{q},\omega)(1-{\tilde{\cal K}}^*(\vc{q},\omega))-
A^*_0(\vc{q},\omega)}
{|{\tilde{\cal K}}(\vc{q},\omega)|^2}
\bigg]
\bigg\}
\end{eqnarray}
\end{widetext}
Equivalently $\sum_{\vc{q},\vc{k'}}
g_\vc{q}(\vc{k_s}-\vc{q},\vc{k'})$ is obtained by 
substituting $\vc{k_s}$ by $\vc{k_s}-\vc{q}$ in (\ref{B3}). 
The domain of integration over $\omega$ can be extended in the complex plane. 
Since the  various reverse dielectric functions in (\ref{reversem}) 
converge to unity when 
$|\omega| \rightarrow \infty$, the integrand goes to zero in this 
limit faster than $1/|\omega|$. As a consequence,  
the integral over $\omega$ can be carried out by closing the 
contour either in the upper half plane or in the lower half plane. 
Since, we assume that the Bose gas is in a stable regime, the poles of 
the dielectric functions have their imaginary part in the lower 
half of the complex plane, while the complex conjugate of these 
functions have their imaginary part in the upper half plane.  
Thus in (\ref{B2}) and in (\ref{B3}), 
contributions in which poles 
only lie in one of these half planes are canceled since, in that case, 
the contour of integration can be chosen in such a way that 
no pole is surrounded.
On the other hand, in order to 
simplify the Eqs. (\ref{B2}) and (\ref{B3}), we can also add  
arbitrary contributions in which poles
lie only in one of these half planes. Taking into account these 
considerations, we arrive at:
\begin{widetext}
\begin{eqnarray}\label{B5}
\sum_{\vc{q},\vc{k'}}(1- \Pi_\vc{k})
g_\vc{q}(\vc{k},\vc{k'})
=
-\sum_{\vc{q}}
\int_{-\infty}^\infty
\frac{d\omega}{2\pi i}
\frac{1}{\omega+i0_+ -
\frac{\vc{k}.\vc{q}}{m}-\frac{\vc{q}^2}{2m}}
\bigg\{
\frac{(n_{\vc{k}}+1)n_{\vc{k}+\vc{q}} 2i{\rm Im}{{\cal K}(\vc{q},\omega)}}
{|{{\cal K}(\vc{q},\omega)}|^2}+ 
\nonumber \\
+
\frac{8\pi a }{mV}(n_{\vc{k}}-n_{\vc{k}+\vc{q}})
\bigg[\frac{2i {\rm Im}A'(\vc{q},\omega)
}{|{{\cal K}(\vc{q},\omega)}|^2}
+
\frac{2i {\rm Im} A_0(\vc{q},\omega)}
{|{\tilde{\cal K}}(\vc{q},\omega)|^2}
\bigg]
\bigg\}
\end{eqnarray}
\begin{eqnarray}\label{B6}
\sum_{\vc{q},\vc{k'}}
g_\vc{q}(\vc{k_s},\vc{k'})
=
-\sum_{\vc{q}}
\int_{-\infty}^\infty
\frac{d\omega}{2\pi i}
\frac{1}{\omega+i0_+ -
\frac{\vc{k_s}.\vc{q}}{m}-\frac{\vc{q}^2}{2m}}
\bigg\{
(n_{\vc{k_s}}+1)n_{\vc{k_s}+\vc{q}}
2i{\rm Im}\left(\frac{{\cal K}(\vc{q},\omega)}
{{\tilde{\cal K}}^*(\vc{q},\omega){\cal K}(\vc{q},\omega)}\right)
\nonumber \\
+
\frac{8\pi a}{mV}(n_{\vc{k_s}}-n_{\vc{k_s+q}})
\bigg[\frac{2i {\rm Im} A'(\vc{q},\omega)}
{{\tilde{{\cal K}}(\vc{q},\omega)}
{{\cal K}^*(\vc{q},\omega)}}
+
(1-\frac{{\cal K}_n(\vc{q},\omega)}{2})
2i{\rm Im}\left(\frac{A_0(\vc{q},\omega)}
{|{\tilde{\cal K}}(\vc{q},\omega)|^2}
\right)
\bigg]
\bigg\}
\end{eqnarray}
\end{widetext}
These expressions can be further simplified if we neglect 
some irrelevant infinitesimal terms proportional to $0_+$. 
Up to these infinitesimal term and using the 
change of variable $\vc{k'}=\vc{k}+\vc{q}$ and the 
formula ${\rm Im}[1/(x+i0_+)]=\pi \delta(x)$,
we notice that for the two different limits 
$n_{\vc{k_s}}/V \rightarrow 0$ or 
$n_{\vc{k_s}}/V$ finite:
\begin{eqnarray}
{\rm Im}{{\cal K}(\vc{q},\omega)}
={\rm Im}
{{\cal K}_n(\vc{q},\omega)}
\nonumber \\ =
\frac{8\pi a }{mV}\sum_{\vc{k'}}
(n'_{\vc{k'-q}}-n'_{\vc{k'}})
\pi\delta(\omega -
\frac{\vc{k'}.\vc{q}}{m}+\frac{\vc{q}^2}{2m})
\end{eqnarray}
\begin{eqnarray}
{\rm Im}\left(\frac{{\cal K}(\vc{q},\omega)}
{{\tilde{\cal K}^*(\vc{q},\omega)}{{\cal K}(\vc{q},\omega)}}\right)
=
\frac{{\rm Im}{\cal K}(\vc{q},\omega)}
{|{\tilde{\cal K}^*(\vc{q},\omega)}{{\cal K}(\vc{q},\omega)}|}
\end{eqnarray}
\begin{eqnarray}
{\rm Im}\left(\frac{A_0(\vc{q},\omega)}
{|{\tilde{\cal K}}(\vc{q},\omega)|^2}\right)
=\frac{ {\rm Im} A_0(\vc{q},\omega)}
{|{\tilde{\cal K}(\vc{q},\omega)}{{\cal K}^*(\vc{q},\omega)}|}
\end{eqnarray}
Finally, it remains to take the imaginary part 
of these expressions, integrate them over $\omega$ and 
plug the results into (\ref{B1}). 
The integration is easy since it involves only delta 
functions. After rearranging terms we notice that, 
in Eq.(\ref{B6}), the term 
proportional to $1-{\cal K}_n(\vc{q},\omega)/2$ 
will not contribute. 
Note that, concerning the term $g_\vc{q}(\vc{k}-\vc{q},\vc{k'})$, 
we must carry out the  change of variable 
$\vc{q} \rightarrow -\vc{q}$ and $\vc{k'}-\vc{q} 
\rightarrow \vc{k'}$.
In this way, we obtain 
Eq.(\ref{K6}) and Eq.(\ref{K7}).

Using (\ref{col34}), a similar reasoning  
allows also to find the expression:
\begin{widetext}
\begin{eqnarray}\label{intg}
\sum_{\vc{k},\vc{k'}} 
g_{\vc{q}}(\vc{k},\vc{k'})=
-
\int_{-\infty}^\infty
\frac{d\omega}{2\pi i}
2i {\rm Im}
\bigg[\frac{1}{{\cal K}(\vc{q},\omega)}
\left(\frac{1}{{\cal K}^*(\vc{q},\omega)}-1\right)
A'(\vc{q},\omega)
+
\frac{1}{{\tilde{\cal K}}(\vc{q},\omega)}
\left(\frac{1}{{\tilde{\cal K}}^*(\vc{q},\omega)}-1\right)
A_0(\vc{q},\omega)\bigg]
\end{eqnarray}
\end{widetext}

\bigskip
\centerline{\bf ACKNOWLEDGMENTS}
PN thanks E. Cornell and  W. Ketterle for a useful discussion about 
superfluidity, N. Cerf for encouraging me to do this work and 
V. Belyi for remarks on the manuscript. 

PN acknowledges financial support from the Communaut\'e Fran\c caise de
Belgique under grant
ARC 00/05-251, from the IUAP programme of the Belgian
government under grant V-18, from the EU under project RESQ
(IST-2001-35759).

\end{document}